\documentclass[preprint,12pt]{article}

\usepackage{graphicx}
\usepackage{amssymb}

\usepackage{lineno}
\usepackage{arxiv}
\usepackage{amsmath}
\usepackage{amsfonts}
\usepackage{subcaption}
\usepackage{physics}

\usepackage{xcolor}
\usepackage[framemethod=TikZ]{mdframed}
\definecolor{ikmgray}{HTML}{5E5E5E}
\definecolor{ikmgreen}{HTML}{C9DA2B}
\newmdenv[frametitle={},
middlelinecolor=ikmgreen,
middlelinewidth=0pt,
backgroundcolor=ikmgray!20,
roundcorner=2pt,
bottomline=false,
leftline=true,
topline=false,
rightline=false,
skipabove=10pt,
skipbelow=10pt,
leftmargin=10pt,
rightmargin=10pt,
innerleftmargin=5pt,
innerrightmargin=10pt,
innertopmargin=10pt,
innerbottommargin=10pt]{Algorithmus}
\usepackage[labelfont=bf, 
format=plain, 
labelsep=endash, 
justification=raggedright 
]{caption}
\DeclareCaptionType{kasten}[Box]
\usepackage{bm}

\usepackage[sort&compress]{natbib}
\DeclareMathOperator*{\argmin}{arg\,min}
\DeclareMathOperator*{\argmax}{arg\,max}
    \DeclareFontFamily{OT1}{pzc}{}
\DeclareFontShape{OT1}{pzc}{m}{it}{<-> s * [1.10] pzcmi7t}{}
\DeclareMathAlphabet{\mathpzc}{OT1}{pzc}{m}{it}
\title{Local approximate Gaussian process regression for data-driven constitutive laws: Development and comparison with neural networks}
\author{
  Jan Niklas Fuhg \\
  Sibley School of Mechanical and Aerospace Engineering \\
  Cornell University, 
   New York, USA \\
     \And
     Michele Marino\\
  Department of Civil Engineering and Computer Science \\
  University of Rome Tor Vergata, \\
  Rome, Italy  \\
   \And
 Nikolaos Bouklas \\
  Sibley School of Mechanical and Aerospace Engineering\\
  Cornell University,
   New York, USA \\
  \texttt{nb589@cornell.edu} \\
  }

\begin{document}
\setlength{\parindent}{0cm}
\maketitle


\begin{abstract}
Hierarchical computational methods for multiscale mechanics such as the $\mathrm{FE}^2$ and FE-FFT methods are generally accompanied by high computational costs. Data-driven approaches are able to speed the process up significantly by enabling to incorporate the effective micromechanical response in macroscale simulations without the need of performing additional computations at each Gauss point explicitly. 
Traditionally artificial neural networks (ANNs) have been the surrogate modeling technique of choice in the solid mechanics community. However they suffer from severe drawbacks due to their parametric nature and suboptimal training and inference properties for the investigated datasets in a three dimensional setting.
These problems can be avoided using local approximate Gaussian process regression (laGPR). This method can allow the prediction of stress outputs at particular strain space locations by training local regression models based on Gaussian processes, using only a subset of the data for each local model, offering better and more reliable accuracy than ANNs.  A modified Newton-Raphson approach is proposed to accommodate for the local nature of the laGPR approximation when solving the global structural problem in a FE setting.
Hence, the presented work offers a complete and general framework enabling multiscale calculations combining a data-driven constitutive prediction using laGPR, and macroscopic calculations using an FE scheme that we test for finite-strain three-dimensional hyperelastic problems. 
\end{abstract}


\section{Introduction}\label{sec::intro}
\begin{figure}
\begin{subfigure}[b]{0.5\linewidth}
\centering
\includegraphics[scale=0.5]{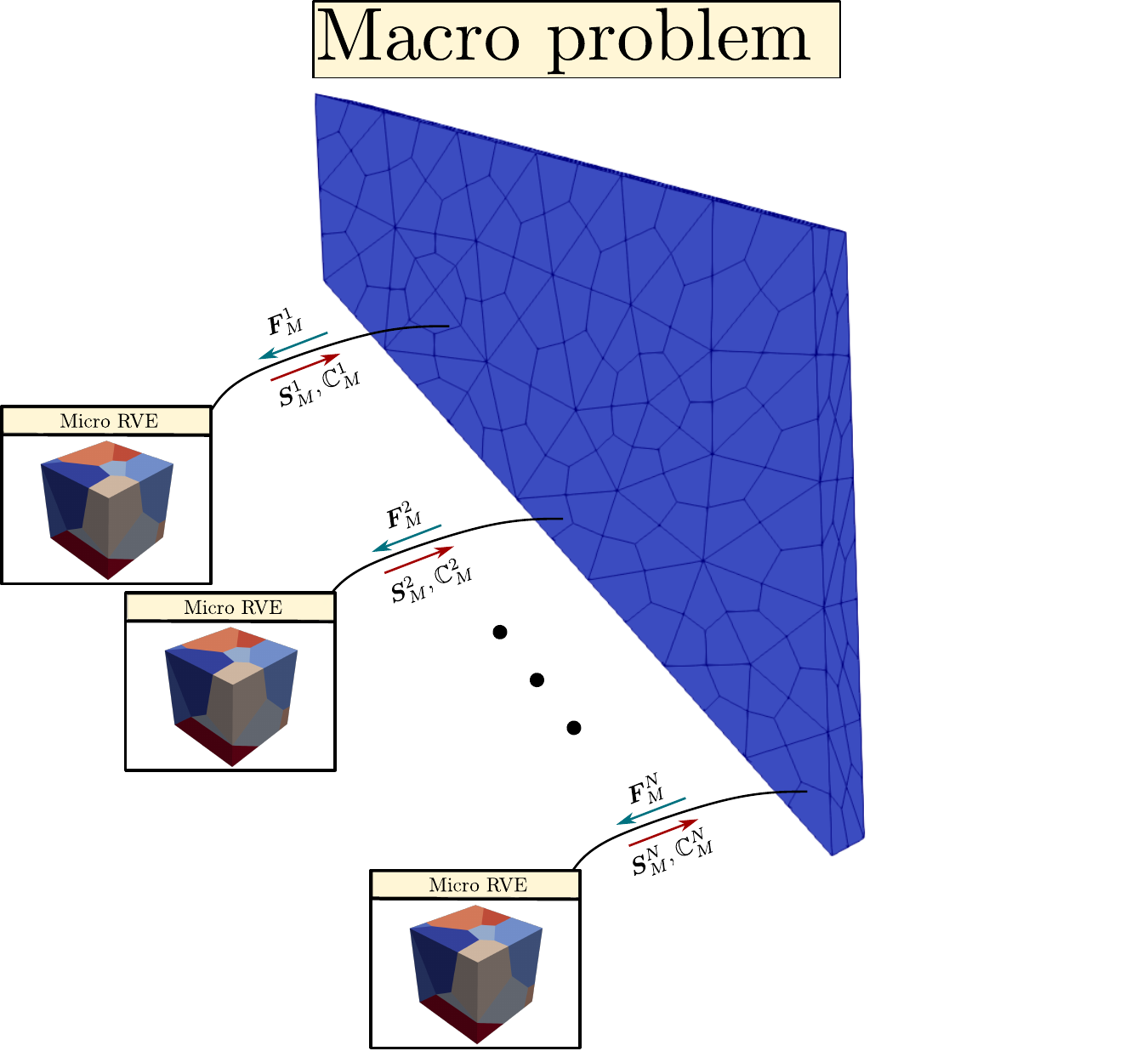}
\caption{Computational homogenization}\label{fig:comHomogenization}
\end{subfigure}%
\begin{subfigure}[b]{0.5\linewidth}
\centering
\includegraphics[scale=0.52]{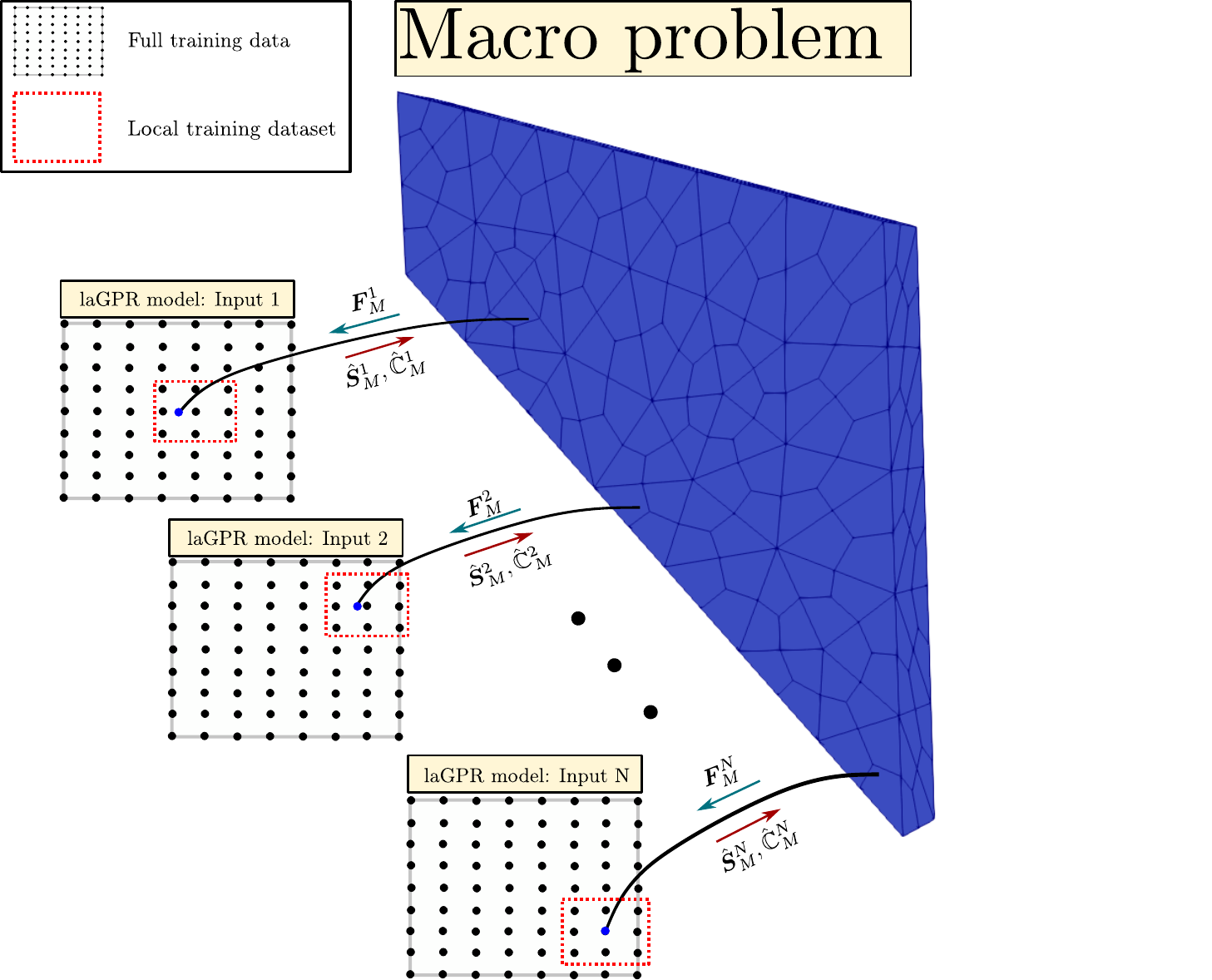}
\caption{laGPR approximation of constitutive response}\label{fig:IdealaGPR}
\end{subfigure}
    \caption{(a) General idea behind computational homogenization schemes. Commonly highly computationally demanding when numerical schemes such as FE$^2$ and FE/FFT are used to obtain microscale response. (b) Local approximate Gaussian process (laGPR) model employed to efficiently approximate the constitutive response. A new model is trained on a reduced subset of the full training dataset based on local proximity to the macroscopic deformation gradient input. }
    \label{fig:comHomogenizationAndLaGPR}
\end{figure}

Machine learning approaches have been an emergent tool in the computational sciences in recent years and have for example recently been shown to reliably solve small and finite strain structural elastomechanics problems \citep{haghighat2021physics, fuhg2021mixed}.
However in solid mechanics, the
mechanics of materials are intrinsically governed by unique phenomena that dictate the response at different length scales.
Hence, the challenge in multiscale mechanics is to identify scale transitions that allow the prediction of the macroscopic properties of the material. These types of calculations necessitate the development of efficient multiscale computational methods.
For a recent overview of homogenization methods and multiscale modeling for nonlinear problems refer to \cite{geers2017homogenization}.
One of the most commonly applied techniques to bridge scales are hierarchical methods, see e.g. \cite{fish2006bridging, fish2010multiscale}, where different length scales are linked in a hierarchical manner. This naturally implies total scale separation and can for example be achieved by computational homogenization, i.e. volume averaging of field variables.
However in order to fully describe complex physical problems the explicit solution in all modeled scales is necessary, leading to an iterative solution process spanning all scales which requires  high computational costs. \\
Computations at the microscale are required due to the lack of a constitutive law that can appropriately capture the response, see Figure \ref{fig:comHomogenization}. This bottleneck can be avoided by using machine learning techniques that aim to learn constitutive laws from data thus eliminating costly online computations at the miscroscale.
On top of their potential to speed up complex and time-consuming simulations data-driven constitutive laws have also gained considerable traction in recent years for their ability to allow for the direct utilization of experimental data.
Continuum scale material models trained with data have been used in
\cite{kirchdoerfer2016data, ibanez2018manifold, gonzalez2019thermodynamically, huang2020machine, ghaderi2020physics} to name a few. \\
In the same context, machine learning techniques have been utilized to facilitate numerical homogenization techniques.
\cite{peng2020multiscale} provide a general overview over possible applications of machine learning techniques in multiscale modeling.
The early works of \cite{yvonnet2009numerically} for small strain two-dimensional elasticity using a multidimensional spline interpolation method have been extended in \cite{le2015computational} to the utilization of neural networks for the same task. Later,
\cite{wang2018multiscale} presented a hybrid approach to bridge scales in poroelastcitiy utilizing recursive neural networks using small strain and two-dimensional examples. 
\cite{liu2019exploring} use a building block system to construct specific material laws based on pre-trained models.  Convolutional neural networks were used in \cite{frankel2019predicting} to predict the effective microstructral response of oligocrystals.
\cite{logarzo373smart} have used Gaussian processes in combinations with recursive neural networks to train the homogenization of a two dimensional RVE with an elastoplastic inclusion in small strain.
A hybrid model-data-driven approach is proposed in \cite{fuhg2021model} to tackle finite-strain hyperelastic problems in plane stress. In that work Gaussian process regression is used as the machine learning formulation to reduce the error between an assumed classical analytical constitutive response and the actual response of the microscale as obtained by finite element (FE) simulations.\\
The presented state-of-the-art formulations are commonly restricted to two-dimensional and small-strain formulations or are exclusively aimed for specific loading paths and constitutive assumptions. 
In this work, we aim to develop a convenient framework to obtain data-driven (model-free) constitutive laws for three-dimensional finite-strain hyperelastic problems at the microscale, and in turn facilitate FE simulations at the macroscopic level. In line with classical approaches in the field, artificial neural networks (ANNs) have been applied to develop surrogate models as a first attempt. However, during the work on this project, we realized that ANNs were only able to generate suboptimal results. This pathology is due to multiple reasons:
\begin{enumerate}
    \item ANNs are parametric models, i.e. they are subject to user-chosen parameter values (number of hidden layers, number of neurons) which significantly influence the prediction quality. These are notoriously chosen by some sort of grid-like search algorithm. However they typically require some experience by the user and are therefore not ideal.
    \item ANNs are regression based approaches which do not exactly represent points of the trained dataset. Furthermore there is no reliable way of fitting certain priors to the neural networks, i.e. requiring it to exactly yield specific output values at a particular input. This property while typically not having much significance for general applications, has negative side effects when training constitutive responses because strain inputs associated with the undeformed configuration do not generate a zero-stress state. This  generally has negative side effects when using the neural network as a complete replacement of the constitutive law in a Finite-Element routine involving Newton-Raphson iterations, i.e. divergence or slow convergence and low accuracy.
    \item ANNs show non-smooth convergence behavior, i.e. as will be shown in this work, adding more unique points to the training dataset does not universally yield a lower error. This property means that ANNs are not reliably trainable and hence hard to evaluate for the user. 
\end{enumerate}
For the above reasons, we shift away from ANNs and instead introduce local approximate Gaussian process regression (laGPR) to obtain data-driven constitutive laws. Based on the early works of \cite{cressie1993statistics} and the later developments by \cite{emery2009kriging} and \cite{gramacy2015local}, the method shows promising results for accurate predictions of general data-dependent mappings. The current work is the first that employs laGPR to predict constitutive responses based on data. It has major advantaged compared to ANNs, for example the method is generally not affected by any of the three problems associated with neural networks as defined above and therefore might prove to be a very useful machine learning technique in the context of data-driven constitutive laws. LaGPR is based on the idea of training models with only a local subset of the whole training dataset, see Figure \ref{fig:IdealaGPR}. Therefore avoiding problems of computational intractability commonly observed with classical Gaussian process regression.
The paper is structured as follows. 
The general problem formulation of computational homogenization of a hyperelastic solid at finite strain  is explained in Section \ref{sec:ProbForm}.
The data-driven concept, consisting of the definition and generation of training data for three dimensional hyperelastic constitutive responses, is given in Section
\ref{sec::DataDrivenConcept}.

The local approximate Gaussian process regression as well as an accompanying modified Newton-Raphson procedure for utilization in a Finite-Element framework accompanied by a short summary of a general neural network formulation is provided in Section
\ref{sec::MachineLearningForm}.
In Section \ref{sec::NumericalExamples}, laGPRs are compared to commonly employed machine learning techniques including neural networks, and are later tested for their use in fitting a high-fidelity constitutive law based on the responses of a three-dimensional representative volume element of a heterogeneous material. In Section \ref{sec::Discussion} the numerical results focusing on the comparison of various machine learning approaches are discussed in the context of constitutive modeling and multiscale analysis.
The paper is concluded in Section
\ref{sec::Conclusion}.

\section{Preliminaries for computational homogenization}\label{sec:ProbForm}
Consider a body $\Omega$ that is microscopically heterogeneous and assume that its microstructure, denoted by the subscript "$m$", is scale-separated from the macroscopic deformation field. Let macroscopic quantities be denoted by subscripts "$M$".
The body motion, in the absence of body forces, is governed by the momentum equation, given by
\begin{equation}\label{eq::Nabla}
\bm{\nabla}_{X,M} \cdot  \left( \bm{F}_{M} \bm{S}_{M} \right) = \rho_{X,M} \ddot{\bm{x}}_{M} 
\end{equation}  
and constrained by initial and boundary conditions. In Eq. \ref{eq::Nabla}, $\bm{S}_{M}$ describes the macroscopic second Piola-Kirchhoff stress tensor, $\bm{F}_{M}$ is the macroscopic deformation gradient, $\rho_{X,M}$ represents the effective density of the reference configuration, $\bm{\nabla}_{X,M}$ is the macroscopic gradient operator with respect to the reference configuration. Furthermore $\bm{x}_{M}$ refers to the position vector in the current configuration, while $\ddot{(\bullet)}$ denotes the second time derivative. 
In order to be able to solve this boundary value problem the necessary constitutive model, relating stress to strain, can be obtained by computational homogenization employing a microstructural representative volume element (RVE). The key concept is shown in Figure \ref{fig:comHomogenization}.  
Consider the body discretized with finite elements. Then, at every integration (e.g. Gauss) point, the macroscopic deformation gradient defines the boundary condition applied to the RVE. With which, in the absence of body forces, the motion of the microscopic scale is governed by the equilibrium equation
\begin{equation}
\bm{\nabla}_{X,m} \cdot \left( \bm{F}_{m} \bm{S}_{m} \right) = \bm{0} .
\end{equation}  
where $\bm{S}_{m}$ represents the microscopic second Piola-Kirchhoff stress tensor, $\bm{F}_{m}$ describes the microscopic deformation gradient and $\bm{\nabla}_{X,m}$ is the microscopic gradient operator with respect to the reference configuration.
Scale separation allows us to define the macroscopic deformation gradient and the first Piola-Kichhoff stress tensors through a volume averaging procedure over the RVE as
\begin{equation}
    \begin{aligned}
    \bm{F}_{M} &= \frac{1}{V_{m}} \int_{V_{m}} \bm{F}_{m} dV_{m}, \\
    \bm{P}_{M} &= \frac{1}{V_{m}} \int_{V_{m}} \bm{P}_{m} dV_{m}.
    \end{aligned}
\end{equation}
The averaged second Piola-Kirchhoff stress tensor can then be computed using
\begin{equation}
    \bm{S}_{M} = \bm{F}_{M}^{-1}  \bm{P}_{M}.
\end{equation}
We then obtain the consistent macroscopic material tangent by
\begin{equation}
        \mathbb{C}_{M} = 2 \frac{\partial \bm{S}_{M}}{\partial \bm{C}_{M}}
\end{equation}
where $\bm{C}_{M} = \bm{F}_{M}^{T} \bm{F}_{M}$ is the macroscopic right Cauchy-Green tensor.
After solving the established microscopic boundary value problem the macroscopic stress tensor $\bm{S}_{M}$ as well as the material tangent $\mathbb{C}_{M}$ can be returned to the respective Gauss point.

\section{Data-driven concept}\label{sec::DataDrivenConcept}
In general, a constitutive model for a homogeneous hyperelastic body is a relation
\begin{equation}
\begin{aligned}
\bm{S} &= \bm{\mathpzc{M}}_{2}(\bm{C}) 
\end{aligned}
\end{equation}
between stress and a (real) second order tensor-valued tensor function $\bm{\mathpzc{M}}_{2}: \mathcal{C} \rightarrow \mathcal{S}$ where $\bm{C} \in \mathcal{C} \subseteq \mathbb{R}^{3 \times 3}$ and $\bm{S} \in \mathcal{S} \subseteq \mathbb{R}^{3 \times 3}$. We call this the major mapping since it yields the essential parameter, which is stress, as an output.
In the context of nonlinear finite element numerical routines it is necessary to access a consistent material tangent as 
\begin{equation}
\begin{aligned}
\mathbb{C} &= \bm{\mathpzc{m}}_{2}(\bm{C}) =  2 \frac{\partial \bm{S}(\bm{C}) }{\partial \bm{C}} .
\end{aligned}
\end{equation}
Hence, we can define a minor tensor-valued mapping $\bm{\mathpzc{m}}_{2}: \mathcal{C} \rightarrow \mathcal{K}$ where $\mathbb{C} \in \mathcal{K} \subseteq \mathbb{R}^{3 \times 3\times 3\times 3}$.
In data-driven constitutive modeling we aim to find approximations for both major and minor mappings by obtaining the metamodels $\hat{\bm{\mathpzc{M}}_{2}}$ of $\bm{\mathpzc{M}}_{2}$ and $\hat{\bm{\mathpzc{m}}_{2}}$ of $\bm{\mathpzc{m}}_{2}$ from a data set consisting of $N$ samples
\begin{equation}\label{def::datasetGeneral}
    \mathcal{D} = \left\lbrace \left( \bm{C}_{i},\bm{S}_{i}, \mathbb{C}_{i} \right) \right\rbrace_{i=1}^{N}\, .
\end{equation}
It needs to be pointed out that neural networks are able to approximate the minor mapping directly from the approximation of the major mapping by automatic differentiation without requiring explicit information about the material tangent (see \cite{huang2020machine}).
However, the dataset of Eq. (\ref{def::datasetGeneral}) represents the most general setting. \\
Even though efforts have been made in the machine learning community to directly learn from second order tensor space mappings (see e.g. \cite{novikov2015tensorizing, yu2018tensor}), the field does not appear advanced enough yet for our purposes. 
Hence, a reconstruction of the original mappings from second order tensor-valued tensor functions ($\bm{\mathpzc{M}}_{2}$ and $\bm{\mathpzc{m}}_{2}$) to a first order tensor mappings ($\bm{\mathpzc{M}}_{1}$ and $\bm{\mathpzc{m}}_{1}$) is necessary.
This can be achieved by observing that the right Cauchy-Green tensor, the second Piola-Kirchhoff stress as well as the material tangent are symmetric tensors which enables us to reduce the number of inputs and outputs by rewriting the second order tensors as vectors and the fourth order elasticity tensor as a second order matrix using the Voigt notation \citep{voigt1928lehrbuch}. 
This leads to the alternative data set
\begin{equation}\label{def::dataset}
    \mathcal{D}_{\text{Voigt}} = \left\lbrace \left( \bm{c}_{\text{i}},\bm{s}_{\text{i}}, \bm{d}_{i}  \right) \right\rbrace_{i=1}^{N}\, .
\end{equation}
where
\begin{equation}
    \begin{aligned}
        \bm{c}_{\text{i}} &= \begin{bmatrix} C_{11} & C_{22} & C_{33} & C_{23} & C_{31} & C_{12} \end{bmatrix}^{T} \in \mathpzc{c} \subseteq \mathbb{R}^{6 \times 1}\\
        \bm{s}_{\text{i}} &=  \begin{bmatrix} S_{11} & S_{22} & S_{33} & S_{23} & S_{31} & S_{12} \end{bmatrix}^{T}\in \mathpzc{s} \subseteq \mathbb{R}^{6 \times 1}
    \end{aligned}
\end{equation}
and $\bm{d}_{i} \in \mathpzc{d} \subseteq  \mathbb{R}^{n \times 1}$ where $n=36$ in the general and $n=21$ in the hyperelastic case.
Hence, the main goal of this paper is to efficiently approximate the major and minor vector-valued mappings $\bm{\mathpzc{M}}_{1}$ and $\bm{\mathpzc{m}}_{1}$ with the surrogates $\hat{\bm{\mathpzc{M}}_{1}}$ and $\hat{\bm{\mathpzc{m}}_{1}}$ respectively.

\subsection{Training points generation}\label{sec::trainingPoints}
At the heart of every data-driven approach lies the training data with its convex hull defining the training region $\mathcal{R}_{tr} \subset \mathpzc{c}$. In the context of approximating surrogate models this region is essential, since it represents the domain in which the output of the surrogate models should be trusted. Due to the highly nonlinear nature of general hyperelastic laws, it is highly unlikely to obtain a general tool for sufficient approximations outside of $\mathcal{R}_{tr}$. Hence, this work does not aim to generate a surrogate model that is valid for values outside of a predefined training domain, but aims for reliable and efficient surrogates inside it.

This section describes the process to generate input data. First of all, physical interpretations of the components of the right Cauchy-Green tensor $\bm{C}_{M}$ are not as intuitive as the equivalent components of the deformation gradient $\bm{F}_{M}$. Hence, efforts were made to define the training region in the deformation gradient space, sample from the components of this tensor, and then translate them to the equivalent $\bm{C}_{M}$ values for the training data.
 However, just randomly sampling from the 9 components of the deformation gradient $\bm{F}$ might lead to nonphysical deformation gradients, i.e. $\det \bm{F} = 0$. In order to enable easier generations of invertible deformation gradients we choose to define $\bm{F} = \bm{F}^{T}$, rendering the applied deformation gradient $\bm{F}_{app}$ to be of the form
    \begin{equation}\label{eq::FappDef}
\begin{aligned}
        \bm{F}_{app} =   \bm{I} &+ F_{12}(\bm{e}_{1} \otimes \bm{E}_{2}+ \bm{e}_{2} \otimes \bm{E}_{1}) + F_{13}(\bm{e}_{1} \otimes \bm{E}_{3}+ \bm{e}_{3} \otimes \bm{E}_{1}) \\&F_{23}(\bm{e}_{2} \otimes \bm{E}_{3}+ \bm{e}_{3} \otimes \bm{E}_{2}) + \sum_{i=1}^{3} (F_{ii}-1) (\bm{e}_{i} \otimes \bm{E}_{i}).
\end{aligned}
\end{equation}
This leaves 6 components $\bm{f}_{app} = \begin{bmatrix} F_{11} & F_{22} & F_{33} & F_{23} & F_{31} & F_{12} \end{bmatrix}^{T}$ to sample from. Relating the components in the form of Eq. (\ref{eq::FappDef}) allows to ensure that $\bm{F}_{app}$  is symmetric positive definite. \\
We observe that a symmetric deformation gradient accounts to $\bm{F} = \bm{U}$, where $\bm{U}$ is a symmetric stretch tensor. 
However this has no effect on the constitutive relationship we are aiming to learn since the input quantity of interest is the right Cauchy-Green tensor and $\bm{C}_{M} = \bm{F}_{M}^{T} \bm{F}_{M} = \bm{U}_{M}^{T} \bm{U}_{M}$. Therefore, for each generated sample in the space of $\bm{f}_{app}$ we can obtain the corresponding stretch $\bm{U}$ which is then applied to the RVE to obtain the stress and material tangent outputs which can be used to train the machine learning tool. \\
The sampling region is defined by the deviation from the unstressed configuration $\bm{f}_{app, 0} = \begin{bmatrix} 1 & 1 & 1 & 0 & 0 & 0 \end{bmatrix}^{T}$. For example, a $10\%$ strain region is given by the six-dimensional hypercube
\begin{equation}
\bm{f}_{app, \pm0.1} = \bm{f}_{app, 0} \pm 0.1 \begin{bmatrix} 1 & 1 & 1 & 1 & 1 & 1 \end{bmatrix}^{T}
\end{equation}
whose convex hull yields the $10\%$ training domain
\begin{equation}
        \mathcal{R}_{tr, 0.1} = \text{conv}(\bm{f}_{app, \pm0.1}).
\end{equation}
Sample points are generated by combining the unstressed configuration with $n_{h}-1$ layers of $3^{6}$ points. Each layer is represented by a six-dimensional hypercube 

For each layer (i.e. hypercube) we place a sample at each vertex position, in the middle of each edge and at the middle of each face of the six-dimensional hypercubes of increasing size growing outwards from the strain value corresponding to the unstressed configuration which guarantees a uniform spread of the sample points as well as a sufficient coverage of samples on the outer surface of the training region.
In order to ensure accurate training around the unstressed configuration, an additional training point at $\bm{f}_{app, 0}$ is added to the dataset. The algorithm is explained in Box \ref{alg::Sampling}.
This approach offers a simple solution to training region extensions by either adding additional sample points by introducing new layers or by stretching existing sample points out by a stretch factor.
\begin{kasten}[ht!]
\begin{Algorithmus}
Input parameters and constants:
Spanned training region $\Delta T$ \\
Number of hypercube layers $n_{h}$ \\
Unstressed applied strain $\bm{f}_{app, 0} = \begin{bmatrix} 1 & 1 & 1 & 0 & 0 & 0 \end{bmatrix}^{T}$
\\
\\
Define incremental cube size $\delta = \frac{\Delta T}{n_{h}}$ \\
For $i=0,1, \cdots , n_{h}-1$:
\begin{itemize}
    \item[] Set hypercube size $\Delta c =(i+1) \delta$ 
    \item[] Sample matrix $\bm{R} = [\bm{f}_{app, -\Delta c}, \bm{f}_{app, 0}, \bm{f}_{app, +\Delta c}]$
    \item[] $\mathcal{D}_{i}$ =$\lbrace$ $3^{6}$ possible combinations of row vectors of $\bm{R}$ $\rbrace$
    \item[] If $i>0$:
    \begin{itemize}
    \item[] Remove $\bm{f}_{app, 0}$ from $\mathcal{D}_{i}$
    \end{itemize}
    \item[] Add $\mathcal{D}_{i}$ to dataset
\end{itemize}
\end{Algorithmus}
\captionof{kasten}{Algorithm for generation of input points based on increasing hypercube sampling.}\label{alg::Sampling}
\end{kasten}

\section{Machine learning formulations}\label{sec::MachineLearningForm}
Given the data set $\mathcal{D}_{\text{Voigt}} = \left\lbrace \left( \bm{c}_{\text{i}},\bm{s}_{\text{i}}, \bm{d}_{i}  \right) \right\rbrace_{i=1}^{N}$ from Eq. (\ref{def::dataset})
consisting of $N$ samples,
the main goal of this paper is to approximate the stress and material tangent vector-valued mappings $\bm{\mathpzc{M}}_{1}$ and $\bm{\mathpzc{m}}_{1}$ with the surrogates $\hat{\bm{\mathpzc{M}}_{1}}$ and $\hat{\bm{\mathpzc{m}}_{1}}$.
Different machine learning approaches could be utilized to train these mappings needed for data-driven constitutive laws. 
Neural networks are the most commonly used approach in recent works \citep{le2015computational, lu2019data, sagiyama2019machine, settgast2020hybrid}. However, as it will be shown within this work,  for a smaller number of samples ($\approx 10,000$) they are unreliable both in terms of accuracy as well as numerical convergence of the finite element scheme they are implemented into. Instead of using neural networks we propose to employ a big data extension of Gaussian process regression as introduced by \cite{kleijnen2020prediction}.
In the following, we provide a short review of neural networks before introducing the concept of Gaussian process regression, and in particular laGPR.

\subsection{Neural networks and Sobolev learning}
A neural network of depth $n_{D}$ consists of one input layer, $n_{D}-1$ hidden layers and one output layer. The $k^{\text{th}}$ hidden layers consists of $n_{k}$ number of neurons. The $(k-1)^{\text{th}}$  has an output $\bm{x}^{k-1} \in \mathbb{R}^{n_{k-1}}$. This output is taken as input to the $k^{\text{th}}$ layer which performs an affine transformation
\begin{equation}\label{eq::NNO}
   \bm{x}^{k} = \mathcal{L} (\bm{x}^{k-1})  = \bm{W}^{k} \bm{x}^{k-1} + \bm{b}^{k}
\end{equation}
where $\bm{W}^{k} $ and $\bm{b}^{k}$ are the weights and biases of the $k^{\text{th}}$ layer. An activation function $\sigma$ is then applied to each component of the output of Eq. (\ref{eq::NNO}) before it becomes the input to the next layer. For regression problems, the activation function of the output layer is chosen as the identity function.
Hence, given an input strain value $\bm{c}$ into the network the output stress $\hat{\bm{s}}$ is computed by
\begin{equation}
    \hat{\bm{s}}(\bm{c}) = (\mathcal{L}_{k} \circ \sigma \circ \mathcal{L}_{k-1}\circ \cdots \circ \sigma \circ \mathcal{L}_{1})(\bm{c})
\end{equation}
where $\circ$ is a composition operator. The tuple of $\bm{\Theta} = \lbrace \bm{W}^{k}, \bm{b}^{k}  \rbrace_{k=1}^{n_{D}}$ defines the trainable parameters of the neural network with optimal values $\bm{\Theta}^{\star}$ that are obtained by an optimization procedure defined over a loss function $L(\bm{\Theta} )$
\begin{equation}
    \bm{\Theta}^{\star} = \argmin_{\bm{\Theta}} L(\bm{\Theta} ).
\end{equation}
The loss function is typically defined as the mean-squared error
\begin{equation}
    L(\bm{\Theta} ) = \text{ } \frac{1}{N} \sum\limits_{i=1}\limits^{N} \norm{ \bm{s}_{i}-\hat{\bm{s}}_{i}}_{2}^{2} .
    \end{equation}

The solution to this minimization problem are commonly defined in an iterative manner by a stochastic gradient descent-style algorithm.
For more information on neural networks, the interested reader is referred to \cite{goodfellow2016deep}.
To the best of the authors knowledge, all approaches to learn constitutive laws from data that are based on neural networks try to generate a surrogate for the major mapping and leave the minor mapping to the backpropagation ability of the network. Some authors, see e.g. \cite{vlassis2020geometric} aim to enhance this formulation with information of the minor mapping by adding a so-called Sobolev term to the loss function
\begin{equation}\label{eq::ANNSOB}
    L_{Sob}(\bm{\Theta} ) = \text{ }  \frac{1}{N} \sum\limits_{i=1}\limits^{N} \underbrace{\norm{ \bm{s}_{i}-\hat{\bm{s}}_{i}}_{2}^{2}}_{\text{Original term}}  + \underbrace{\lambda \norm{\frac{\partial \bm{s}_{i}}{\partial \bm{c}_{i}} - \frac{\partial \hat{\bm{s}}_{i}}{\partial \bm{c}_{i}}}_{2}^{2}}_{\text{Optional: Sobolev term}}
\end{equation}
where $\lambda\geq 0$ can either be set by the user or can be chosen as a trainable parameter.

\begin{figure}
    \centering
    \includegraphics[scale=0.6]{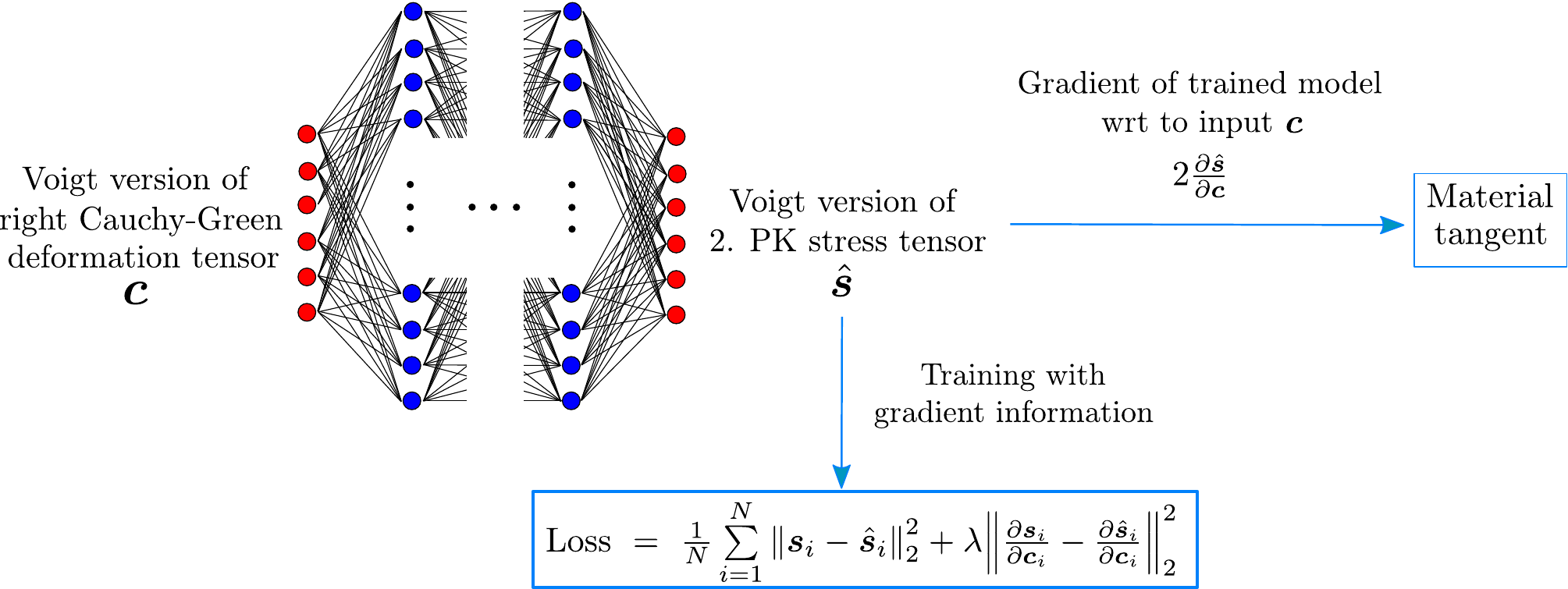}
    \caption{Schematic of the training procedure with Sobolev training.}
    \label{fig:my_label}
\end{figure}

\subsection{Gaussian process regression and local approximate Gaussian process regression}
Gaussian process regression (GPR) is based on the idea that outputs of mappings are more correlated the closer their inputs are in space. The technique has been developed and has been popular for decades in geostatistics, where it is known as Kriging \citep{matheron1963principles} and in the computer experiments community where they are called surrogate models \citep{sacks1989design}. Only recently have they been employed more as a prediction tool in the machine learning literature \citep{rasmussen2003gaussian}.
They are also a common choice for defining adaptively sampled design of experiments \citep{fuhg2020state}.
Their popularity is due to multiple reasons, as they perform proficiently for out-of-sample model predictions and furthermore their nonparametric structure allows for analytical capabilities not seen in other commonly utilized machine learning techniques \citep{gramacy2016lagp}.\\
Before exploring GPR, let the dataset of Eq. (\ref{def::dataset}) be redefined by combining the stress and tangent outputs, i.e. major and minor mapping outputs, into a single output vector $\bm{y} \in \mathbb{R}^{42 \times 1}$, yielding the new dataset tuple
\begin{equation}\label{def::datasetGPR}
    \mathcal{D}_{\text{Voigt}, GPR} = \left\lbrace \left( \bm{c}_{\text{i}}, \bm{y}_{i}=\lbrace \bm{s}_{\text{i}}, \bm{d}_{i} \rbrace  \right) \right\rbrace_{i=1}^{N}\, .
\end{equation}
Since the training dataset consists of $N$ data points, we can rewrite the whole output data as a $42N \times 1$ stacked vector of outputs $\bm{ y}^{tp}$, i.e.
\begin{equation}
\bm{ y}^{tp}=\begin{bmatrix} {\bf y}_{1} \\ \vdots \\ {\bf y}_{N} \end{bmatrix}\, .
\end{equation}

 
Consider the deterministic output to be a realization of a stochastic process of the form:
\begin{equation}\label{eq:Gauss_general}
\bm{ Y}(\bm{ c}) = \bm{\mu} + \bm{ A} \bm{ Z}\, ,
\end{equation}
with $\bm{ Y}$ the $42$-dimensional output vector, $\bm{\mu}$ the $42$-dimensional vector representing the mean of the Gaussian process, $\bm{ A}$ a $(42\times 42)$ positive-definite matrix (the first set of unknown parameters), and $\bm{ Z}$ a $42$-dimensional vector of mutually independent Gaussian processes \citep{svenson2010multiobjective}. 
In GPR metamodeling, an autocorrelation function $R$ between two inputs $\bm{ c}$ and $\bm{ c}'$ has to be chosen, describing the rate of correlation decay as a function of the distance between the two inputs. In the following, we consider it to be fixed as a Mat\'{e}rn 3/2 function  of the form \citep{matern1960spatial}:
\begin{equation}\label{eq:autocorr}
\begin{aligned}
R(\bm{ c}, \bm{ c}', \bm{\theta}_{i})  =  \prod_{k=1}^{6}  \left( 1 + \dfrac{\sqrt{3} \abs{c_{k} - c'_{k}}}{\theta_{i,k}} \right) \exp \left(-\dfrac{\sqrt{3} \abs{c_{k} - c'_{k}} }{\theta_{i,k}}  \right) \, \text{,}
\end{aligned}
\end{equation}
where $c_k$ and $c'_k$ are the $k$-th components of inputs ${\bf c}$ and ${\bf c}'$ (with $k=1,\ldots,6$), and $\boldsymbol{\theta}=\left[ \boldsymbol{\theta}_{1}, \ldots, \boldsymbol{\theta}_{42} \right]^{T}$ is a vector where each $\boldsymbol{\theta}_{i}$ collects $6$ hyperparameters $\theta_{i,1}, \ldots, \theta_{i,6}$ (the second set of unknown parameters). The covariance of ${\bf Y}$ between two values is given by
\begin{equation}
Cov({\bf Y}({\bf c}), {\bf Y}({\bf{c}}')) = {\bf A} {\bf R}({\bf c},{\bf c}'){\bf A}^{T}
\end{equation}
where ${\bf R} \in \mathbb{R}^{42\times 42}$ reads
\begin{equation}
{\bf R}({\bf c},{\bf c}') = \text{diag}\lbrace R({\bf c}, {\bf c}', \boldsymbol{\theta}_{1}), \cdots  , R({\bf c}, {\bf c}', \boldsymbol{\theta}_{42}) \rbrace\, .
\end{equation}
In case we encounter two equivalent input strains, the matrix can be defined by
\begin{equation}
{\bf \Sigma}_{0}({\bf c}) = Cov({\bf Y}({\bf c}), {\bf Y}({\bf{c}})).
\end{equation}
Consider the covariance matrix $\boldsymbol{\Sigma}$ of ${\bf Y}$ as a $(42n_{tp} \times 42n_{tp})$ covariance matrix given block-component-wise by
\begin{equation}
[\boldsymbol{\Sigma}]_{ij} = \begin{cases}
{\bf \Sigma}_{0}({\bf c}_i) & \text{for } i=j\, , \\
Cov({\bf Y}({\bf c}_{i}), {\bf Y}({\bf{c}}_{j})) & \text{else}\, .
\end{cases}
%
\end{equation}
Then, the output of GPR at a point belonging in the convex-hull of the training dataset ${\bf c}_{\star} \in \mathpzc{c}$, not necessarily belonging to the training dataset (i.e., ${\bf c}_{\star} \neq {\bf c}_1, \ldots , {\bf c}_{n_{tr}}$), can be approximated by
\begin{equation}\label{eq:mean}
\begin{aligned}
\hat{{\bf y}}({\bf c}_{\star}) =  \hat{\bm{\mu}} + \boldsymbol{{\Pi}}({\bf c}_{\star}) \boldsymbol{\Sigma} \big({\bf y}^{tp}- {\bf \mathcal{F}} \hat{\bm{\mu}} \big),
\end{aligned}
\end{equation}
where $\boldsymbol{{\Pi}} \in \mathbb{R}^{42 \times 42 n_{tp}}$ and $\hat{\bm{\mu}} \in \mathbb{R}^{42}$ are given by: 
\begin{subequations}
\begin{align}
& \boldsymbol{{\Pi}}({\bf c}_{\star}) = \begin{bmatrix}
Cov({\bf Y}({\bf c}_{\star}), {\bf Y}({\bf c}_{1})) & \cdots & Cov({\bf Y}({\bf c}_{\star}), {\bf Y}({\bf c}_{n_{tr}})) \end{bmatrix}\, ,\\
& \hat{\bm{\mu}}= ({\bf \mathcal{F}}^{T} \boldsymbol{\Sigma}^{-1} {\bf \mathcal{F}})^{-1} {\bf \mathcal{F}}^{T} \boldsymbol{\Sigma}^{-1} {\bf y}^{tp}\, ,
\end{align}
\end{subequations}
with ${\bf{\mathcal{F}}} = \bm{1}_{n_{tp}} \otimes {\bf{I}}_{42}$ obtained from  a $n_{tp}$-dimensional vector of ones $\bm{1}_{n_{tp}}$ and a $(42 \times 42)$ unit matrix ${\bf{I}}_{42}$ through the Kronecker operator $\otimes$ yielding ${\bf{\mathcal{F}}} \in \mathbb{R}^{42 n_{tp} \times 42}$.

It should be noted that points belonging to the training dataset are exactly interpolated. However, the output prediction of the GPR model still depends on the tuple of unknown parameters ${\bf A}$ and ${\bm{\theta}}$, which need to be determined by a "training" process. 
However, when we consider ${\bf A}$ to be a unit matrix, the outputs are essentially uncorrelated, which will be assumed in the following.
The remaining unknown values of ${\bm{\theta}}$ can then be estimated using a restricted maximum likelihood approach, given by:
\begin{equation} \label{eq:optim_ML}
\begin{aligned}
\hat{{\bm{\theta}}} = \argmax_{ {\bm{\theta}}^{\star}} &\left[-\frac{1}{4} \log(|\bm{\Sigma}|) \log({\bf \mathcal{F}}^{T} \bm{\Sigma}^{-1} {\bf \mathcal{F}}) + \right. \\
&\left.-\frac{1}{2} ({\bf y}^{tp} - {\bf \mathcal{F}} \hat{\bm{\mu}})^{T} \bm{\Sigma}^{-1} ({\bf y}^{tp} - {\bf \mathcal{F}} \hat{\bm{\mu}}) \right]\, .
\end{aligned}
\end{equation}
After solving this optimization problem, a set of parameters is obtained that yields the GPR surrogate model   describing the training dataset the best (in a generalized least-squared setting).  
The surrogate can then be used to make predictions using equation (\ref{eq:mean}). 
\subsubsection*{Local approximate Gaussian process regression} 
Traditionally, GPR as described above, suffers from computational intractability when dealing with large datasets since training and inference scale with $N^3$ as $N$ is the number of sample points (see Eqs. (\ref{eq:mean}) and (\ref{eq:optim_ML})). Using this approach directly for data-driven constitutive modeling, which might require a multiple of $10,000$ samples, is not feasible. Different numerical methods have been proposed to alleviate this problem by e.g. using approximation techniques, leading to the so-called scalable Gaussian process regression (sGPR) \citep{deisenroth2015distributed, wilson2015thoughts, wilson2015kernel}. Others use graphical processing units (GPUs) to make GPR utilizable for large datasets \citep{franey2012short}.

Additionally, when using sGPR we encounter a drawback in the context of constitutive law approximations, since sGPR in contrast to GPR is no longer exactly fitting points belonging to the training dataset. Therefore there are no guarantees that the undeformed configuration yields a zero stress output. This property is crucial when using data-driven constitutive modeling in FE modeling at the macroscopic level, since it allows for an accurate and converging Newton-Raphson procedure.

Therefore we consider an alternative approach, called local approximate Gaussian process regression (laGPR). It can be considered as a modern version of local Kriging, a formulation employed by
\cite{cressie1993statistics} for geostatistics applications.
It employs the following idea: approximate the mapping output at a particular strain input $\bm{c}_{\star}$, by only using a subset of $n$ samples of the whole dataset $\mathcal{D}_{n, \text{Voigt}, GPR}(\bm{c}) \subseteq \mathcal{D}_{\text{Voigt}, GPR}$ of equation (\ref{def::datasetGPR}). Hereby, the sub-design generally consists of inducing points $\mathcal{X}_{n}$ close to $\bm{c}_{\star}$. The concept behind this approach is that, when looking at the autocorrelation function of Eq. (\ref{eq:autocorr}), the correlation between input points of $\mathcal{D}_{\text{Voigt}, GPR}$ decays quickly for strain input values $\bm{c}' \notin \mathcal{X}_{n}$ which are far away from $\bm{c}_{\star}$, i.e. the points $\bm{c}'$ have a diminishing influence on the prediction of the model at point $\bm{c}_{\star}$. Therefore, ignoring values which are far away from the particular input $\bm{c}_{\star}$ will allow us to work with a much smaller $n$-sized matrices instead of the original matrices of size $N$.
For a visualization of the decay of the Mat\'{e}rn 3/2 as well as other autocorrelation functions, the interested reader is referred to \cite{fuhg2019adaptive}. \\
Different variations of formulations for the inducing point set $\mathcal{X}_{n}$ have been explored and tested in the literature
\citep{emery2009kriging, gramacy2015local, datta2016hierarchical, gramacy2016lagp}.
\cite{kleijnen2020prediction} propose to select $n$ points of the full dataset $N$ that are closest to an input $\bm{c}_{\star}$ and use these samples to build a local approximating Gaussian process regression model.
The proximity between two points in the input space is defined by the euclidean distance
\begin{equation}
    d(\bm{c}_{\star}, \bm{c}) = \sqrt{(\bm{c}_{\star}- \bm{c})^{T} (\bm{c}_{\star}- \bm{c})}.
\end{equation}

With the laGPR approach we are able to benefit from all the advantages that GPR offers, while making use of the large datasets that are required to accurately approximate three dimensional finite strain hyperleastic constitutive laws. 
However, the laGPR approach has some other drawbacks when using it as a prediction tool for constitutive laws in an FE setting.  Constantly changing the metamodel at each Gauss point, in turn also results in small constitutive response deviations between evaluations of the residual and the consistent tangent through Newton-Raphson iterations which leads convergence issues. 
In order to generate a consistently converging scheme, we propose a modified Newton-Raphson approach specific to the laGPR, in which the surrogate model at a Gauss-point is "frozen" if the Frobenius norm of the residual right Cauchy-Green tensor $\norm{\Delta \bm{C}^{g}}_{\mathcal{F}}$ is smaller than a threshold value $C_{tol}$. This threshold value needs to be chosen in accordance to the covered space of the training points at each local surrogate model. The modified Newton-Raphson algorithm is summarized in Box \ref{alg::mNR}.
\begin{kasten}[ht!]
\begin{Algorithmus}
Initial values $\bm{u}_{0} = \bm{u_{k}}$\\
Loadstep $\overline{\lambda}$\\
$\bm{C}$-space tolerance $C_{tol}$\\
Convergence tolerance $G_{tol}$\\
\\
Iteration loop $i=0,1, \cdots$ until convergence
\begin{itemize}
\item[] At each Gauss point $g$:
\begin{itemize}
\item[] Obtain right Cauchy-Green tensor residual norm $ \norm{\Delta \bm{C}^{g}}_{\mathcal{F}}$
\item[] If $ \norm{\Delta \bm{C}^{g}}_{\mathcal{F}}$ $> C_{tol}$:
\begin{itemize}
\item[] Generate local surrogate $\hat{\mathcal{M}}_{i}^{g}$ via laGPR
\end{itemize}
\item[] else:
\begin{itemize}
\item[] Set $\hat{\mathcal{M}}_{i}^{g}$ $=$ $\hat{\mathcal{M}}_{i-1}^{g}$
\end{itemize}
\item[] From $\hat{\mathcal{M}}_{i}^{g}$ obtain local stress and tangent contributions
\end{itemize}
\item[] Assemble global residual $\bm{G}(\bm{u_{i}}, \overline{\lambda})$ and tangent matrix $\bm{K}_{T}(\bm{u_{i}})$
\item[] Compute the increments $\bm{K}_{T}(\bm{u_{i}}) \Delta (\bm{u_{i+1}}) = - \bm{G}(\bm{u_{i}}, \overline{\lambda})$
\item[] Update primary variable  $\bm{u}_{i+1} = \bm{u}_{i} + \Delta \bm{u_{i+1}}$
\item[] Test for convergence
\begin{equation*}
 \begin{cases}
 \norm{\bm{G}(\bm{u_{i}}, \overline{\lambda})}_{2}
  \leq G_{tol}: & \text{Set } \bm{u_{k+1}} = \bm{u_{i+1}} \text{ and STOP} \\
   \norm{\bm{G}(\bm{u_{i}}, \overline{\lambda})}_{2}
 > G_{tol}: & \text{Set } i = i+1 \text{ and CONTINUE} 
  \end{cases}
\end{equation*}
\end{itemize}
\end{Algorithmus}
\captionof{kasten}{Modified Netwon-Raphson approach for local approximate Gaussian process regression.}\label{alg::mNR}
\end{kasten}

\clearpage
\section{Numerical experiments}\label{sec::NumericalExamples}
In this section we test the proficiency and applicability of laGPR for data-driven constitutive modeling. Additionally, we test the applicability of laGPR 
along with the proposed modified Newton-Raphson scheme, for structural FE modeling. The suggested setup, which follows the offline-online paradigm, is  proposed as a computationally efficient alternative to hierarchical methods for numerical homogenization. 
In Section \ref{sec::Comparison}, we compare the predictive capabilities of laGPR to other machine learning formulations, and in particular to artifical neural networks,  given an analytical expression for a constitutive law following a transversely isotropic Neo-Hookean response.
Hereafter, in Section \ref{sec::Homogenizatoin}, laGPR is utilized to fit a microstructural constitutive response obtained from numerical, FE-based, homogenization of a three dimensional RVE composed of 12 inclusions with periodic boundary conditions, and used to solve multiple  structural FE problems using the proposed modified Newton-Raphson scheme.

\subsection{Comparison of laGPR with other machine learning techniques}\label{sec::Comparison}
\begin{figure}
    \centering
    \includegraphics[scale=0.5]{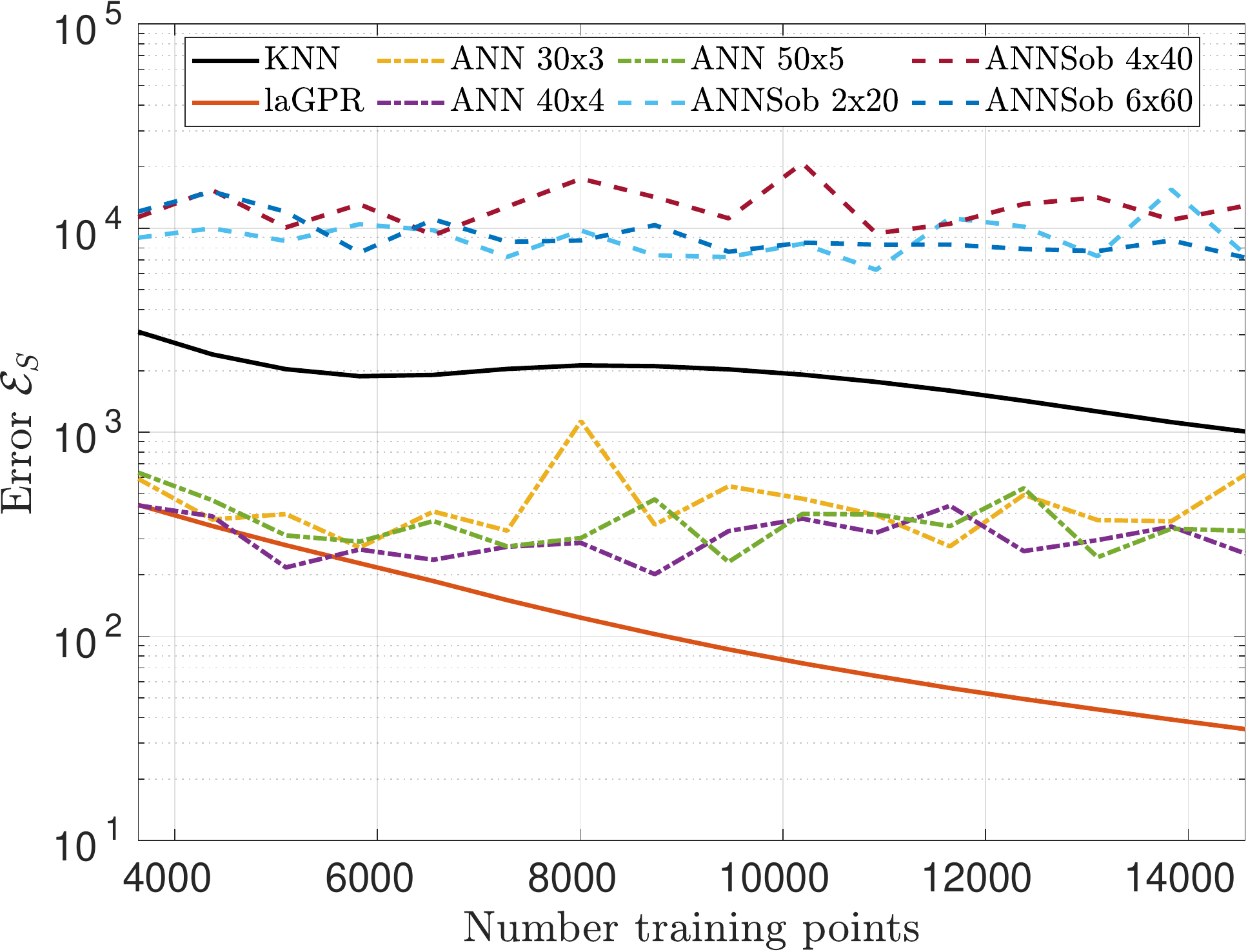}
    \caption{Error $\mathcal{E}_{S}$ in \ref{eq::ErrorS}: Comparison with respect to 10000 reference points (placed with LHD sampling) between laGPR, k-nearest neighbors and variations of ANN with and without the Sobolev term (hereon termed ANNSob and ANN equivalently). The Sobolev weight is chosen to be $\lambda=0.01$, which after an initial study was found to yield the best results. The terms $i \times j$ denote networks with $i$ hidden layers a $j$ neurons. }
    \label{fig:ErrorBetweenMethods}
\end{figure}

\begin{figure}
\begin{subfigure}[b]{.5\linewidth}
\centering
    \centering
    \includegraphics[scale=0.3]{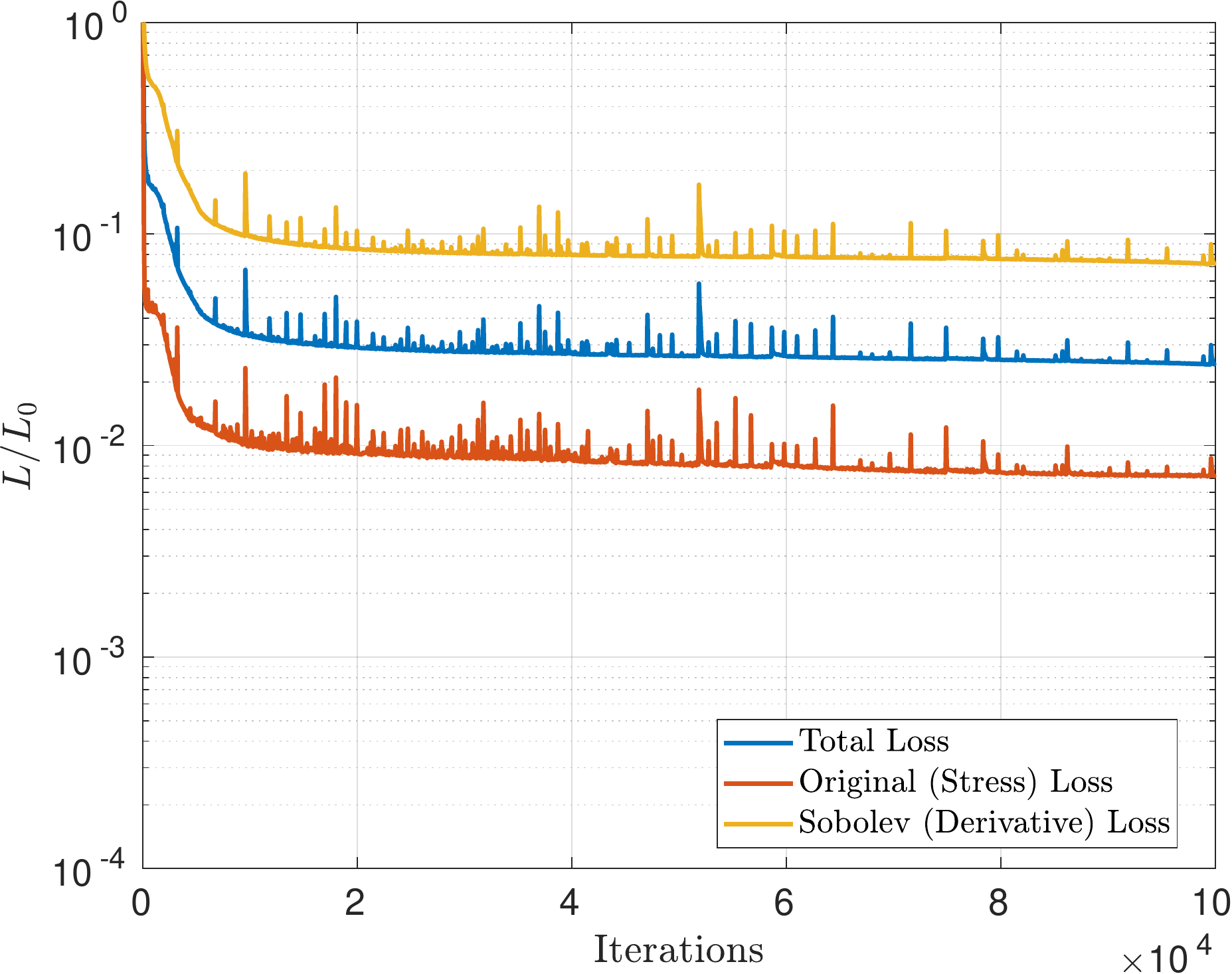}
    \caption{ANNSob 6x60 ($\lambda = 0.01$)}\label{}
\end{subfigure}%
\begin{subfigure}[b]{.5\linewidth}
\centering
    \centering
    \includegraphics[scale=0.3]{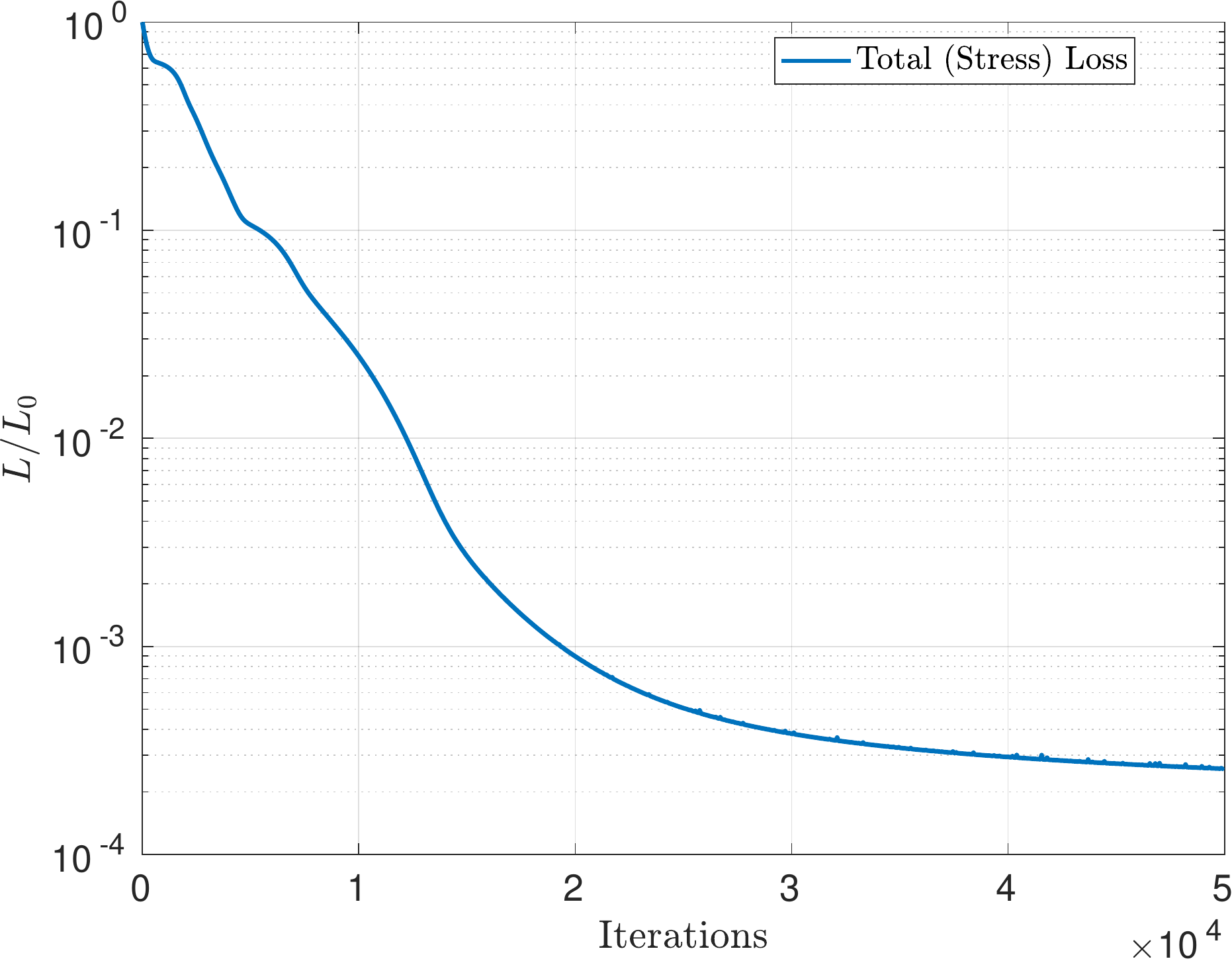}
    \caption{ANN 5x50 ($\lambda = 0.0$)}\label{}
\end{subfigure}%
    \caption{Loss $L$ normalized by initial loss value $L_{0}$. Loss evolution of neural networks training with and without Sobolev term ($\lambda = 0.01$) with 14561 points.}
    \label{fig:ANNLossEvolution}
\end{figure}

\begin{figure}
\begin{subfigure}[b]{0.5\linewidth}
\centering
\includegraphics[scale=0.32]{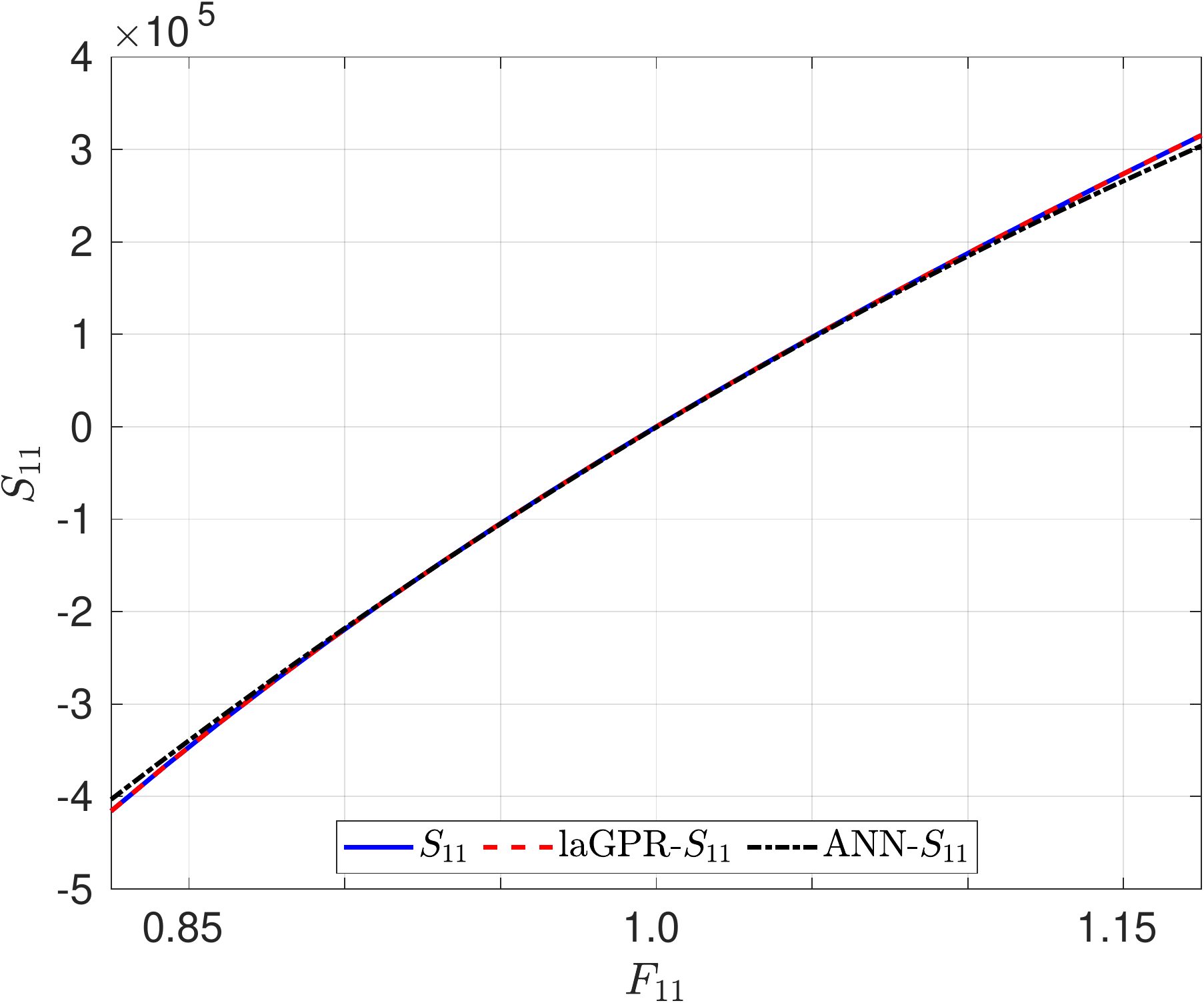} 
\caption{}\label{}
\end{subfigure}%
\begin{subfigure}[b]{0.5\linewidth}
\centering
\includegraphics[scale=0.32]{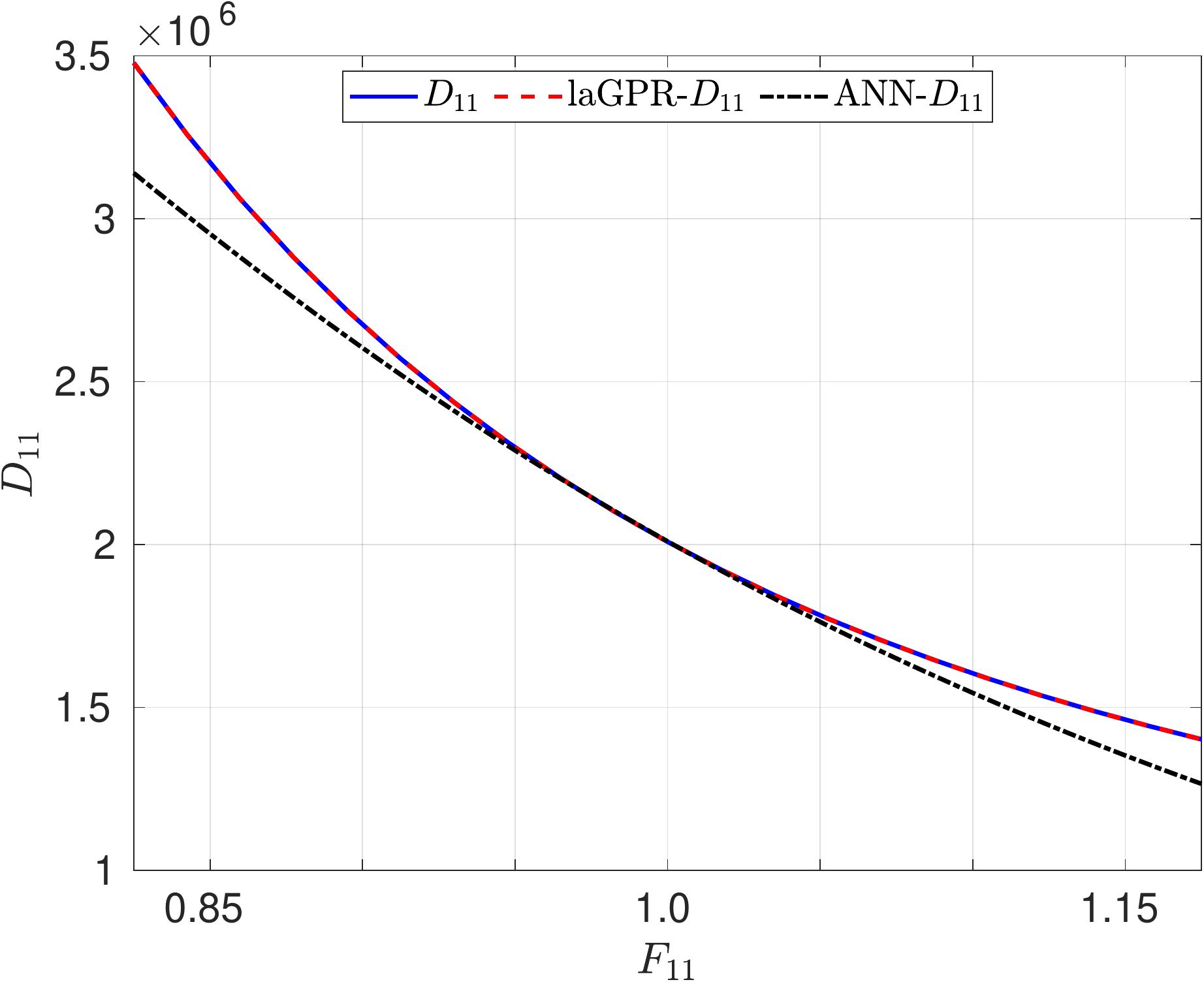}
\caption{}
\end{subfigure}

\begin{subfigure}[b]{0.5\linewidth}
\centering
\includegraphics[scale=0.32]{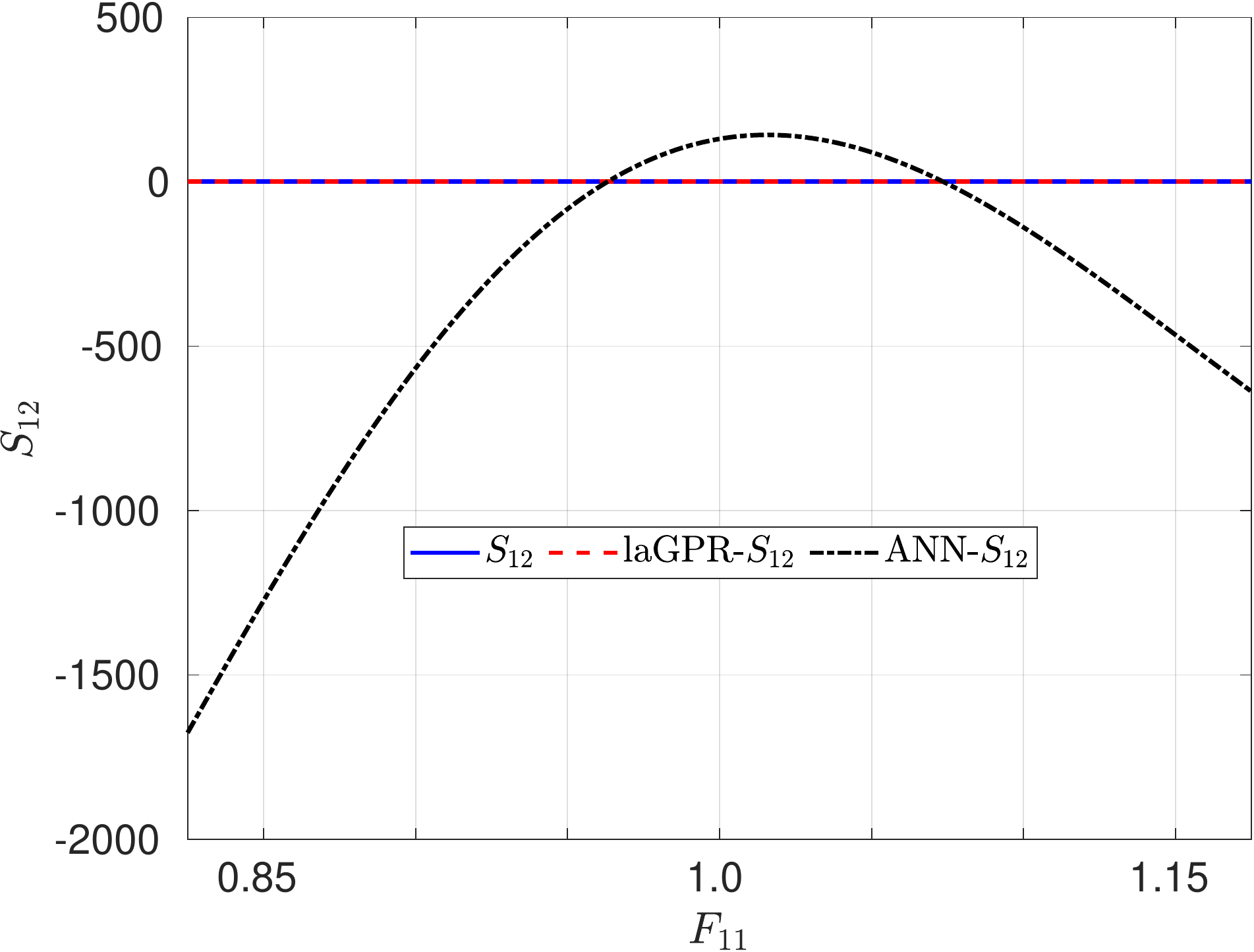} 
\caption{}\label{}
\end{subfigure}%
\begin{subfigure}[b]{0.5\linewidth}
\centering
\includegraphics[scale=0.32]{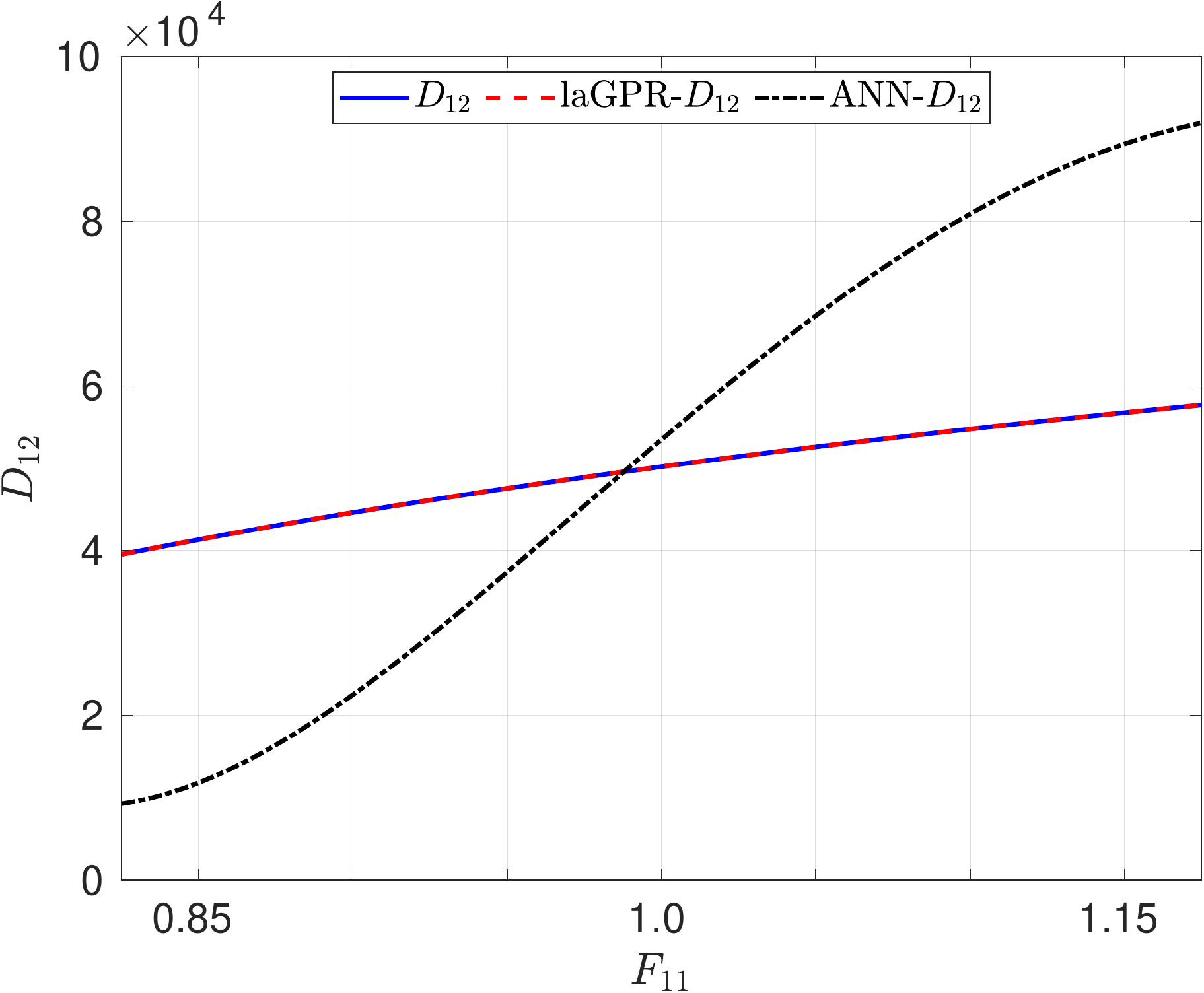} 
\caption{}\label{}
\end{subfigure}
\caption{Analytical constitutive responses (blue) vs. laGPR (red) and ANN (black) predictions defined over a range of deformation gradient components $F_{11}$ (corresponding to uniaxial loading path in the direction 1). LaGPR regression model and ANN with 5 hidden layers a 50 neurons trained with 14651 samples respectively.}\label{fig::AnalyActualCurves}
\end{figure}

\begin{figure}[h!]
\begin{subfigure}[b]{0.5\linewidth}
\centering
\includegraphics[scale=0.32]{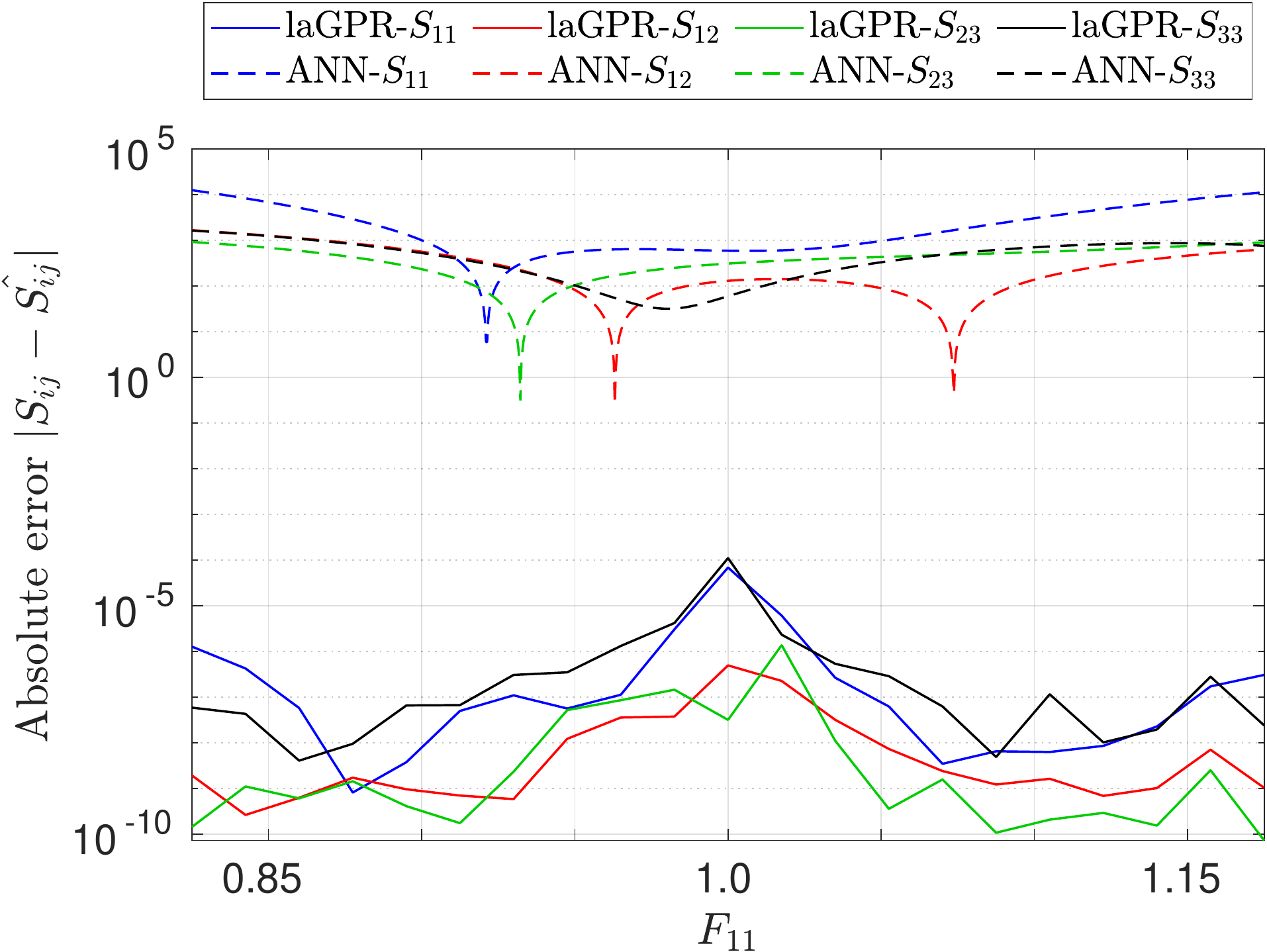} 
\caption{Uniaxial loading in $F_{11}$}\label{fig::AnalyticalF11}
\end{subfigure}%
\begin{subfigure}[b]{0.5\linewidth}
\centering
\includegraphics[scale=0.32]{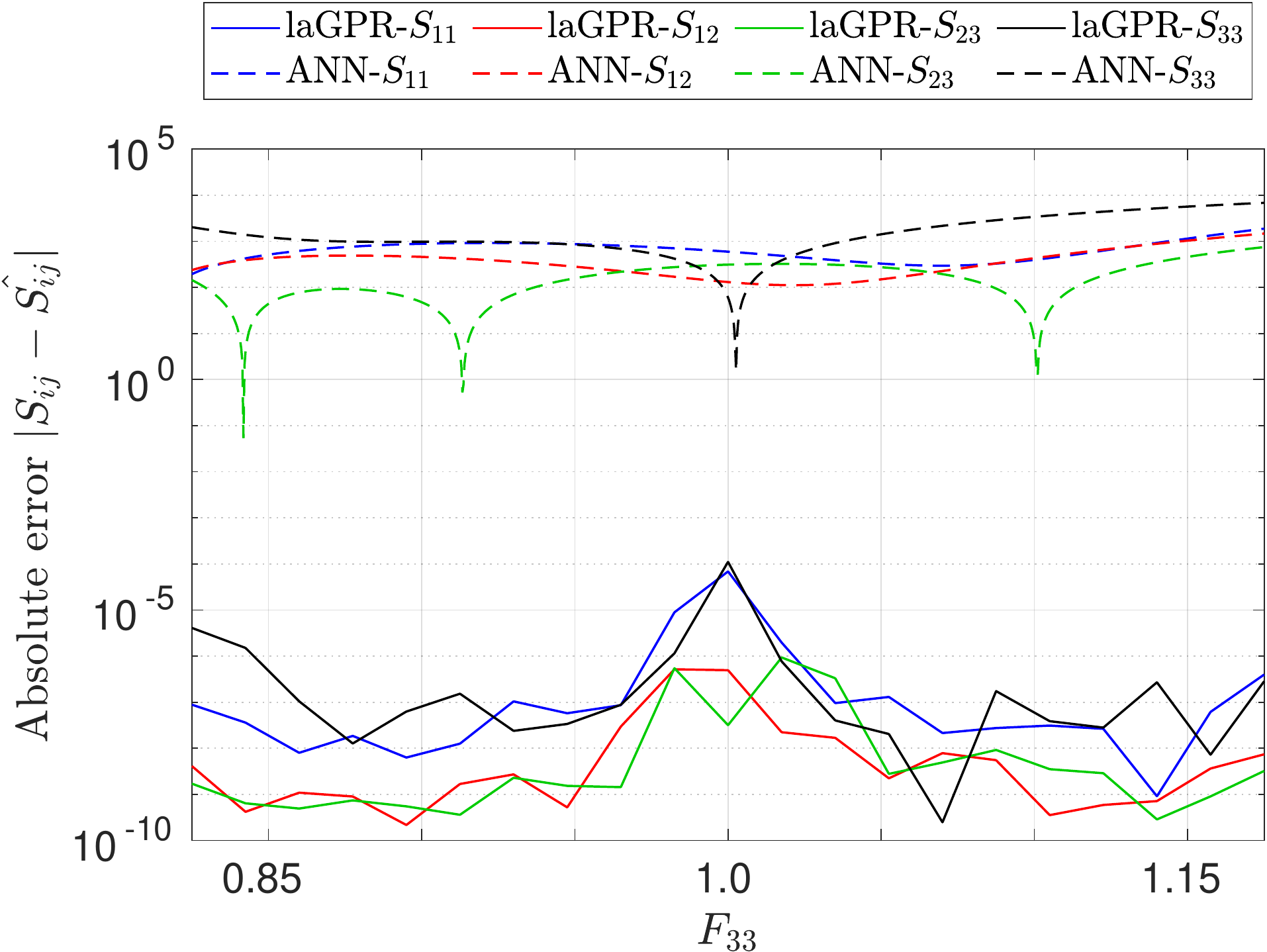}
\caption{Uniaxial loading in $F_{33}$}
\end{subfigure}

\begin{subfigure}[b]{1.0\linewidth}
\centering
\includegraphics[scale=0.32]{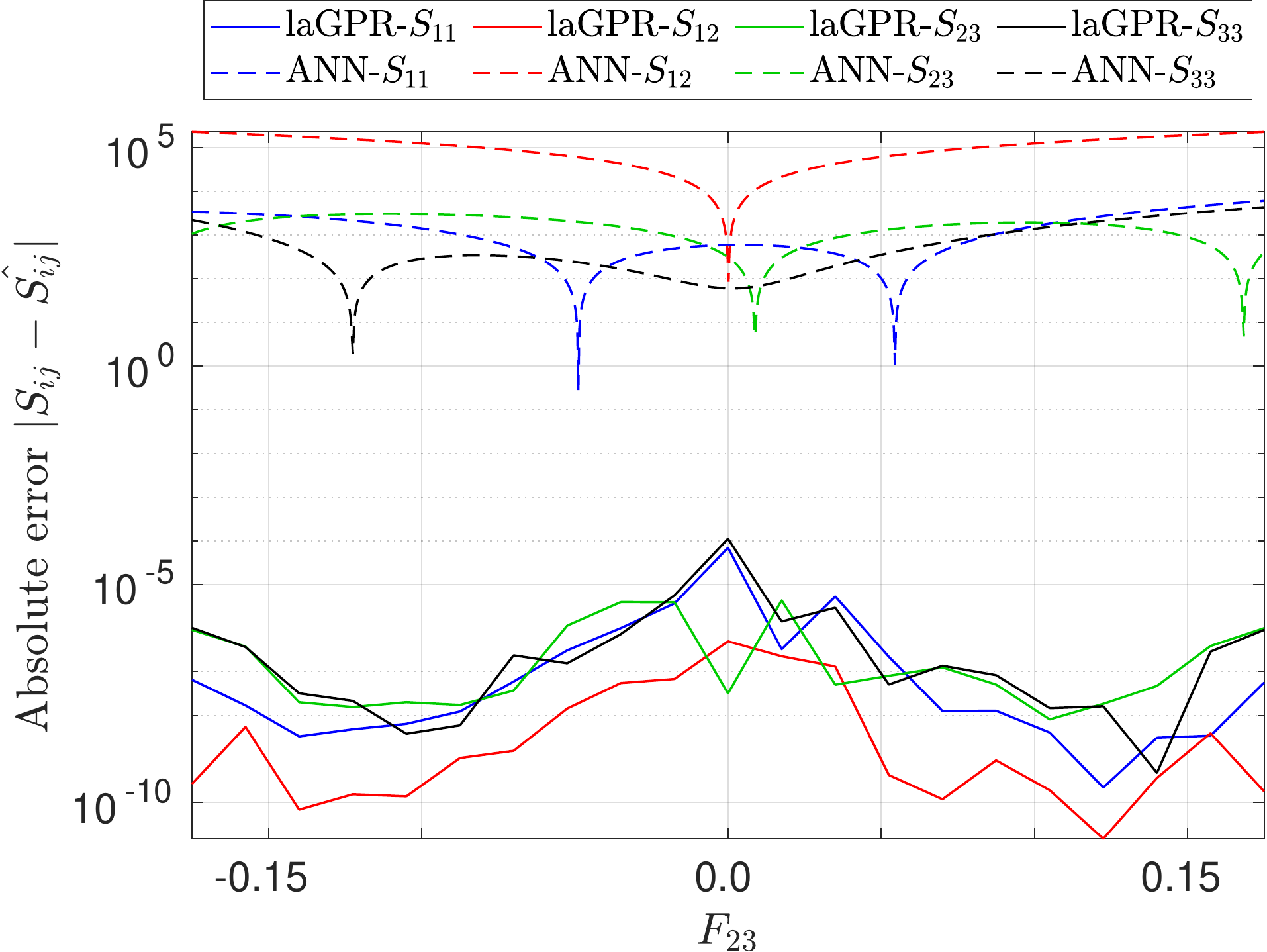} 
\caption{Uniaxial loading in $F_{23}$}\label{}
\end{subfigure}
\caption{Absolute error to analytical constitutive response for three uniaxial load cases ($F_{11}$, $F_{33}$ and $F_{23}$). laGPR regression model and ANN with 5 hidden layers a 50 neurons trained with 14651 samples respectively.}\label{fig::AnalylaGPRandGroundTruth}
\end{figure}

In this section we compare laGPR to neural network as well as k-nearest neighbor (kNN) regression using 1 neighbor, see e.g. \cite{altman1992introduction}, in their effectiveness of reproducing the constitutive response obtained with an analytical transversely isotropic hyperelastic law given by
\begin{equation}
    \Psi = 0.5 \mu (\text{tr}(\bm{C}) - 3- 2 \log (J)) + 0.5 \beta (J-1)^{2} + 0.5 \gamma (I_{4}-1)^{2},
\end{equation}
which yields the stress
\begin{equation}
    \bm{S} = \mu (\bm{I} - \bm{C}^{-1}) + \beta J (J-1) \bm{C}^{-1} +2 \gamma (I_{4}-1) \bm{a}_{0} \otimes \bm{a}_{0},
\end{equation}
where $\bm{a}_{0}= [1,0,0]^{T}$ is a unit vector representing a fiber reinforced direction, $I_{4} = \bm{C}: \bm{a}_{0} \otimes \bm{a}_{0}$ the fourth principal invariant, and normalized shear, bulk and fiber reinforcement moduli of $\mu=6.175e5$, $\beta=5e4$ and $\gamma=1.8e5$. \\
The neural network training is performed using Pytorch \citep{NEURIPS2019_9015}, the activation function is chosen as the commonly applied Rectified Linear Unit (ReLU). Furthermore we employ the well-known and commonly used Adam's optimizer \citep{kingma2014adam} to train the networks with a learning rate of $10^{-4}$.
The KNN regression algorithm from scikit \citep{scikitlearn} was used.
The laGPR was implemented from scratch in Python \footnote{Codes will be made public after acceptance of this paper.}. However, implementations exist also in R-packages \citep{gramacy2016lagp}. We use $n=100$ points for finding the inducing points. The hyperparameters were optimized using a Hooke-Jeeves-type (pattern search) algorithm \citep{kelley1999iterative}, commonly employed in the constraint optimization of GPR hyperparameters, see e.g \cite{lophaven2002dace}.   \\
Without loss of generality, we consider any trained data-driven constitutive law to be useful in a $17.5\%$ training domain. Hence, we will spread the training points in this region depending on a certain number of layers $n_{h}$ according to the procedure outlined in Box \ref{alg::Sampling}.
For testing, we also randomly sample $N_{t} =10,000$ points in the training domain using Latin Hypercube sampling \citep{mckay2000comparison}.
With reference to these points we define the mean stress output error as
\begin{equation}\label{eq::ErrorS}
    \mathcal{E}_{S} = \frac{1}{6} \sum_{i=1}^{6} \sum_{j=1}^{N_{t}} \abs{\hat{\bm{s}}_{i}^{j} - \bm{s}_{i}^{j}}^{2}
\end{equation}
where $\bm{s}_{i}^{j}$ represents the $j$-th (Voigt) stress component of the $i$-th sample point, and $\hat{\bm{s}}_{i}^{j}$ its trained counterpart.
Figure \ref{fig:ErrorBetweenMethods} plots this error value over an increasing number of points in the training dataset for kNN, laGRP and different variations of the hyperparameters for ANNs.
Furthermore, training the ANNs with ($\lambda=0.01$) and without ($\lambda=0$) the Sobolev term of Eq. (\ref{eq::ANNSOB}) is tested.
In order to avoid vanishing or exploding gradient problems with the ANNs, the input and output training data sets were normalized into the range $[0,1]$ before training.\\
Multiple observations can be drawn from the error results of Figure \ref{fig:ErrorBetweenMethods}.
Firstly, it can be seen that, by adding new points, laGPR yields by far the lowest error value for a higher number of training points. Secondly, both neural network formulations are not consistently able to reduce the error even with a higher number of training points available. This is a major disadvantage when using neural networks as data-driven constitutive models since from a user-perspective more information should generally always lead to a better fitted response. It can be highlighted that this problem is not occuring when fitting the data and making predictions using laGPR. Here, an increase in the training data consistently produces a better surrogate. The KNN formulation also converges with increasing data points, however its error is significantly larger than laGPR. \\
Lastly, there is a major difference in the quality of the trained neural networks when a Sobolev term is added to the loss function. The error value of the networks trained with derivatives information is even worse than using a simple kNN approach. This observation can be be understood by looking at the loss values over the training process.
The loss evolutions over the training iterations for 14651 training points for a neural network with 6 hidden layers a 60 neurons trained with a Sobolev term in the loss function and a neural network with 5 hidden layers a 50 neurons (which was arbitrarily chosen from the range of trained models) are depicted in Figure \ref{fig:ANNLossEvolution}. Similar results were found for the trained models with different choices of parameters.
It can be seen that for these training sets neural networks with Sobolev contribution appear not to be able to reduce the error as without it, even when the contribution is chosen to have a small effect, with $\lambda=0.01$.
Even though for the  ANN with no Sobolev training the loss appears to show proficient convergence with a smooth error reduction pointing towards no overfitting issues, the overall error in comparison to laGPR is still significantly worse (Figure \ref{fig:ErrorBetweenMethods}). In light of this and under consideration that all neural networks were trained with the same learning rate and optimization scheme, the training procedure of ANNs seems unreliable, i.e. an increase in training points and/or more trainable parameters does not seem to necessarily lead to a decrease in error.
The differences between laGPR and neural networks can be understood from a different perspective by comparing the direct constitutive response outputs with the analytical solution, see Figure \ref{fig::AnalyActualCurves}. Here, considering a uniaxial load case corresponding to a variation of $F_{11}$ in the applied deformation gradient (see Eq. \ref{eq::FappDef}), the predicted $S_{11}$ and $S_{12}$ stress values as well as two predicted components of the material tangent $D_{11}$ and $D_{12}$ are compared to the analytical solution. It can be seen that laGPR fits all responses with a very high accuracy, whereas neural networks show worse prediction results. However, it can be noted that the neural network prediction is comparatively accurate for the stress output $S_{11}$ for this load case, whereas constant smaller stress output values such as $S_{12}$ are significantly harder to predict precisely for a neural network. This is due to the fact that ANNs are optimizing the trainable parameters based on the $L_2$ error-norm, which promotes the reduction of magnitude-wise larger output values since they tend to produce larger errors.
However, the output difference between the two machine learning formulation are highlighted more profoundly when looking at absolute errors between the predictions and the analytical solutions for the uniaxial loading cases along the three coordinate axes, see  
Figure \ref{fig::AnalylaGPRandGroundTruth}. It can be seen that laGPR show significantly more accurate performance by reaching a factor of $10^{10}$ difference for some input cases.

\begin{figure}[h!]
\begin{subfigure}[b]{0.33\linewidth}
\centering
\includegraphics[scale=0.45]{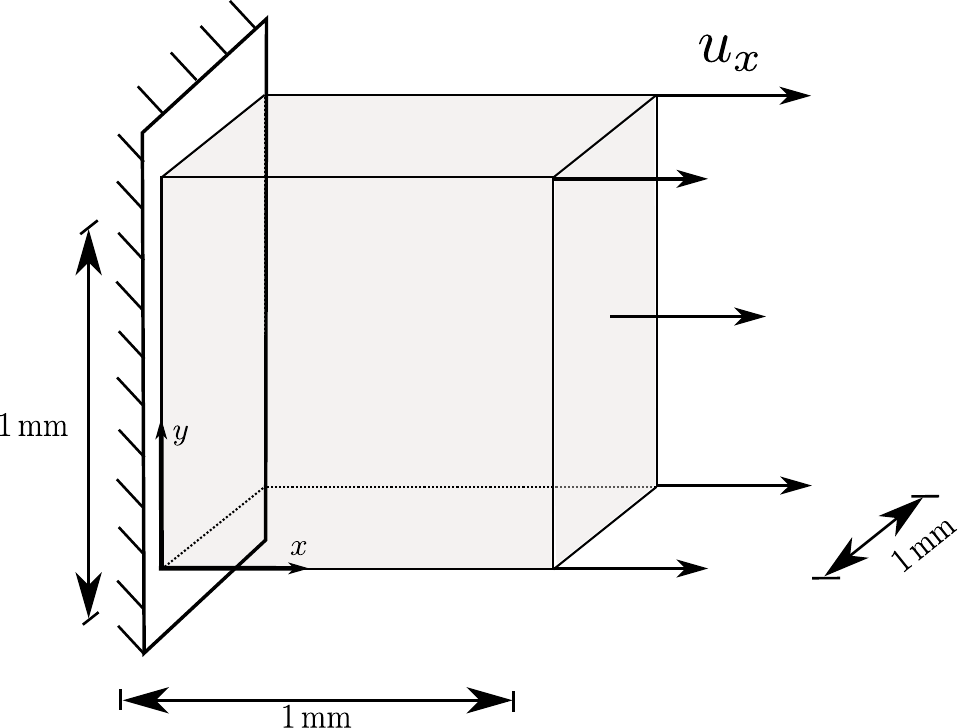} 
\caption{}\label{}
\end{subfigure}%
\begin{subfigure}[b]{0.33\linewidth}
\centering
\includegraphics[scale=0.45]{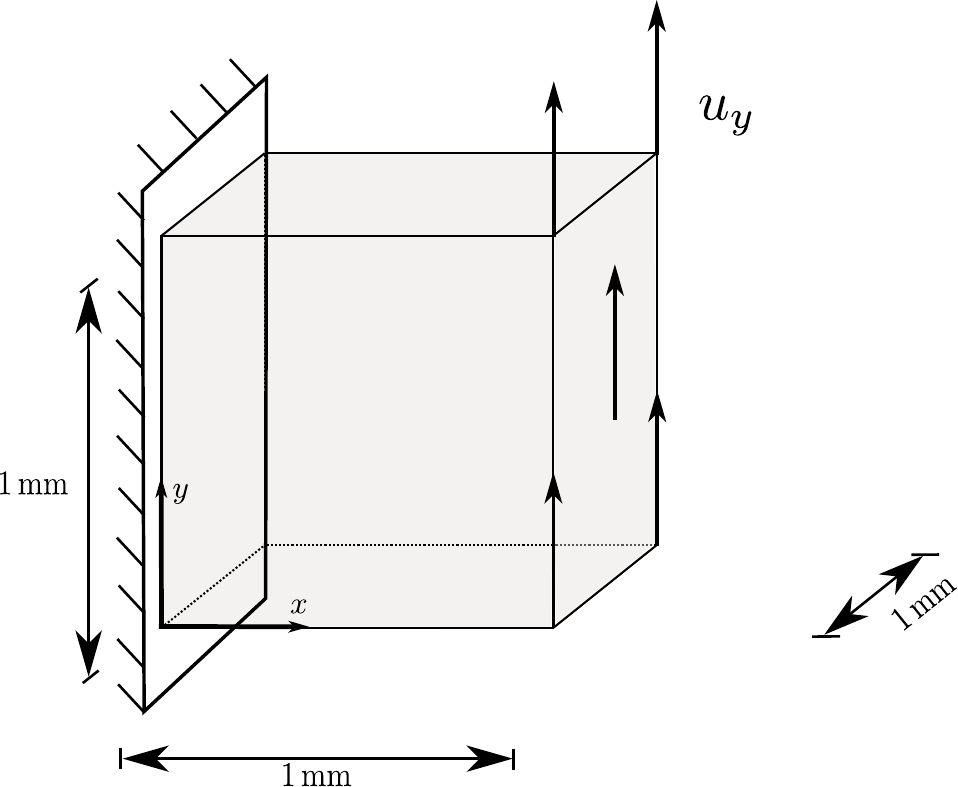}
\caption{}
\end{subfigure}
\begin{subfigure}[b]{0.33\linewidth}
\centering
\includegraphics[scale=0.45]{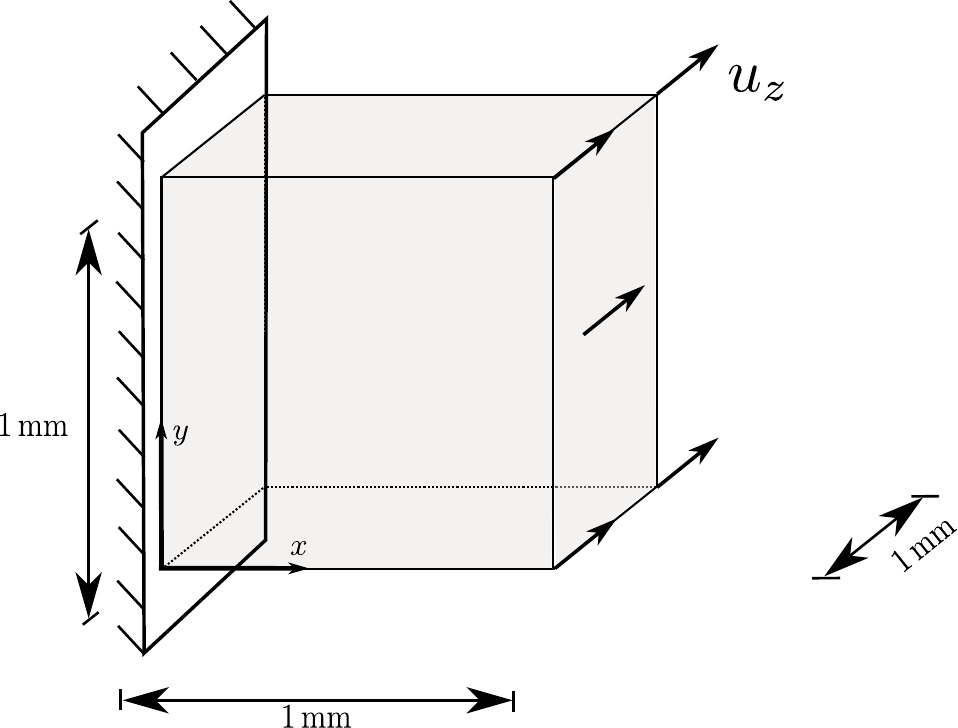} 
\caption{}\label{}
\end{subfigure}
\caption{Three structural FE application problems to study the convergence rates of the investigated machine learning techniques. A clamped cube, discretized with $8x8x8$ elements, is loaded by means of displacememt driven loading cases. Applied displacements are chosen such that a maximum absolute deformation gradient component value of $15\%$ is reached. }\label{fig::ClampedCube}
\end{figure}

\begin{figure}[h!]
\begin{subfigure}[b]{.5\linewidth}
\centering
\includegraphics[scale=0.33]{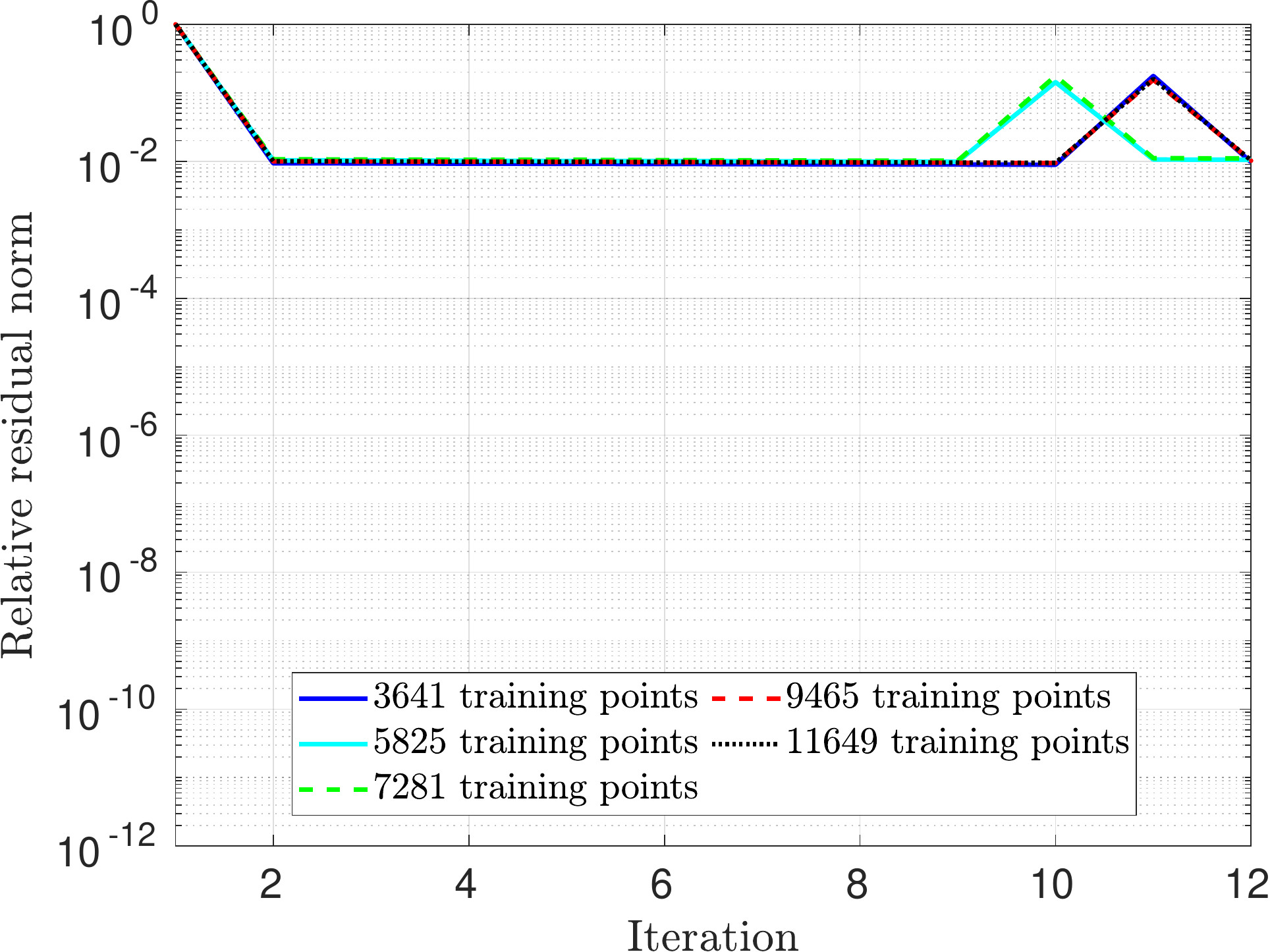} 
\caption{KNN}\label{}
\end{subfigure}%
\begin{subfigure}[b]{.5\linewidth}
\centering
\includegraphics[scale=0.33]{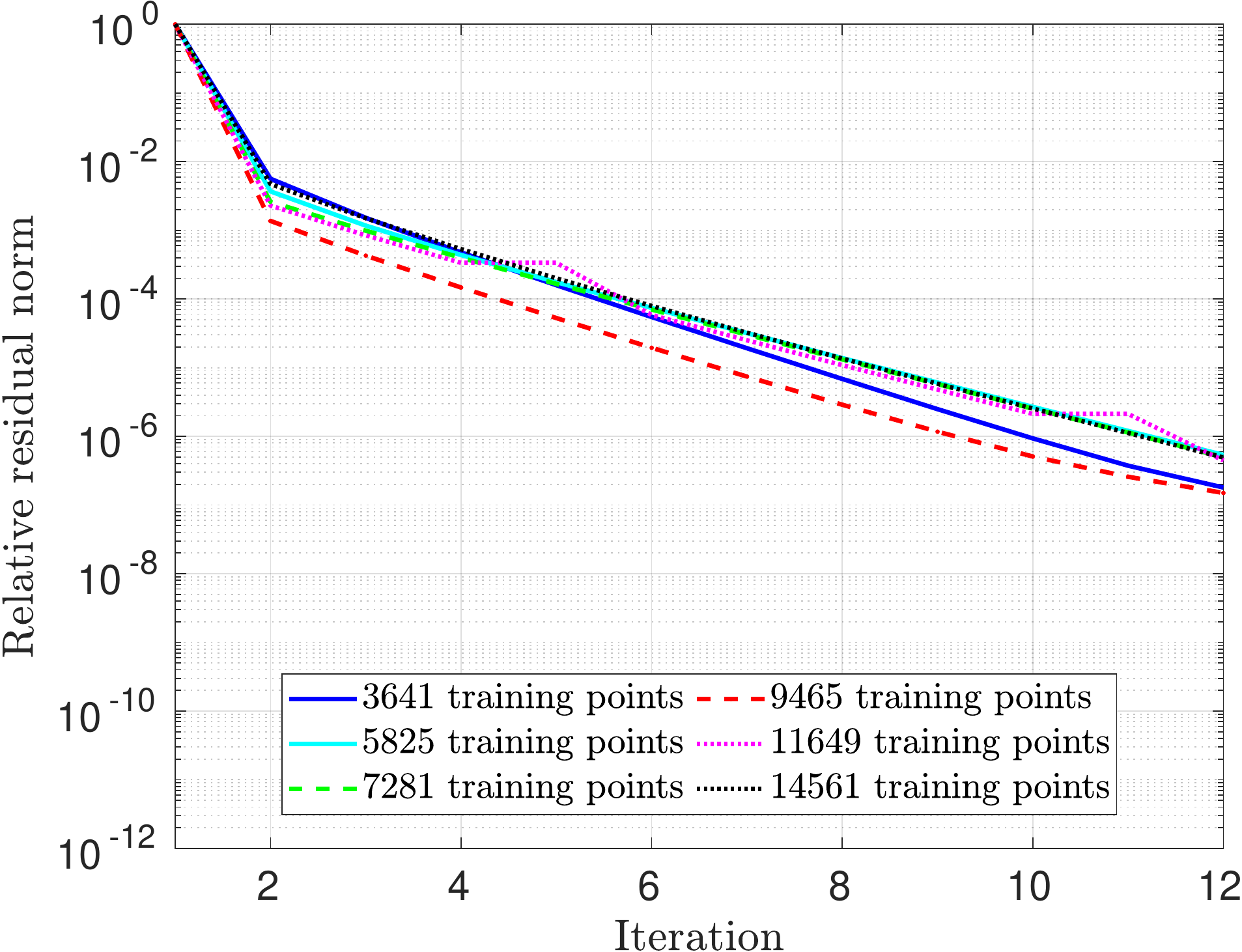}
\caption{ANN 5x50}\label{fig::ConvergenceANN}
\end{subfigure}

\begin{subfigure}[b]{.5\linewidth}
\centering
\includegraphics[scale=0.33]{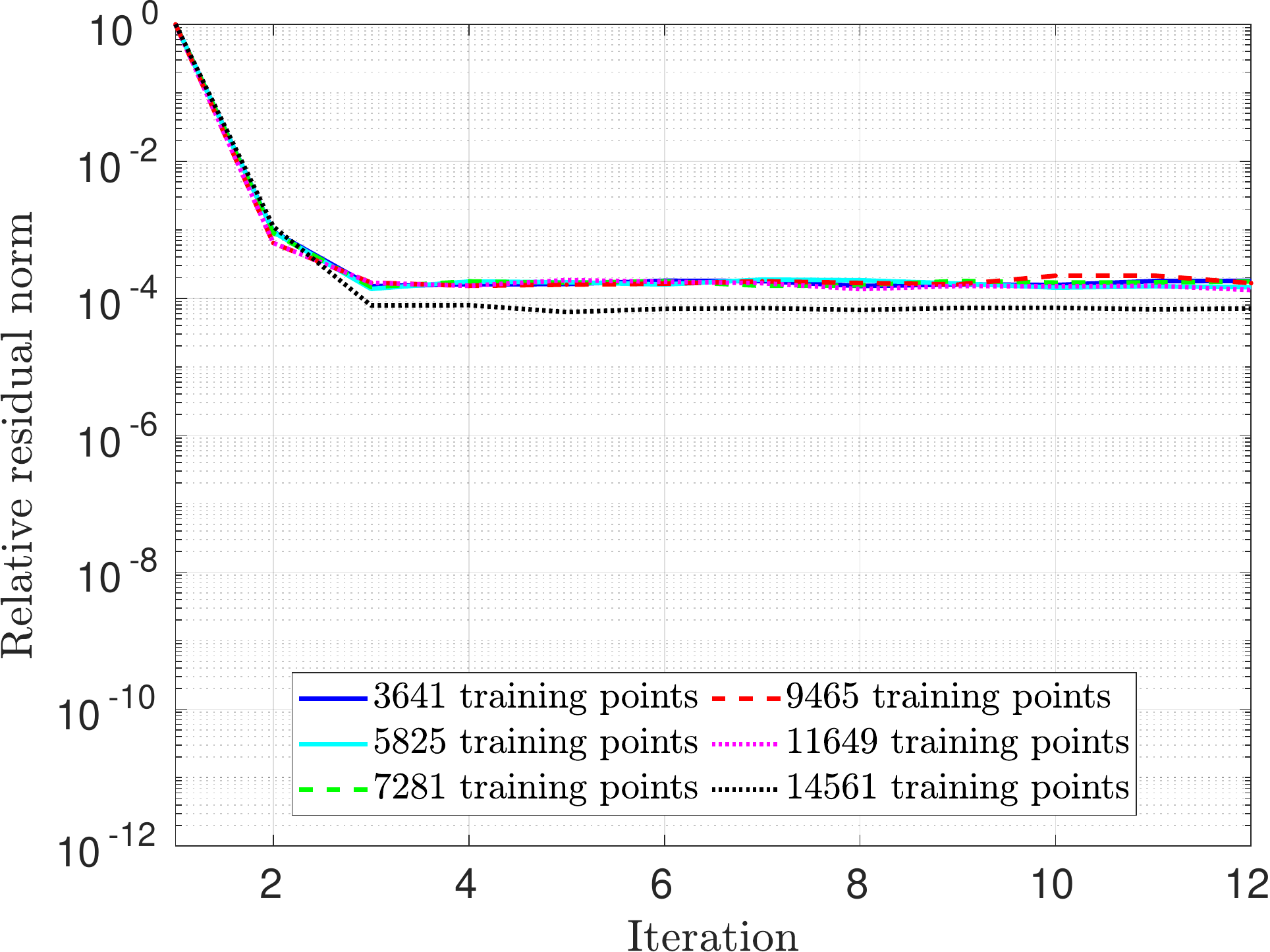} 
\caption{laGPR - No freeze}\label{fig::noFreeze}
\end{subfigure}%
\begin{subfigure}[b]{.5\linewidth}
\centering
\includegraphics[scale=0.33]{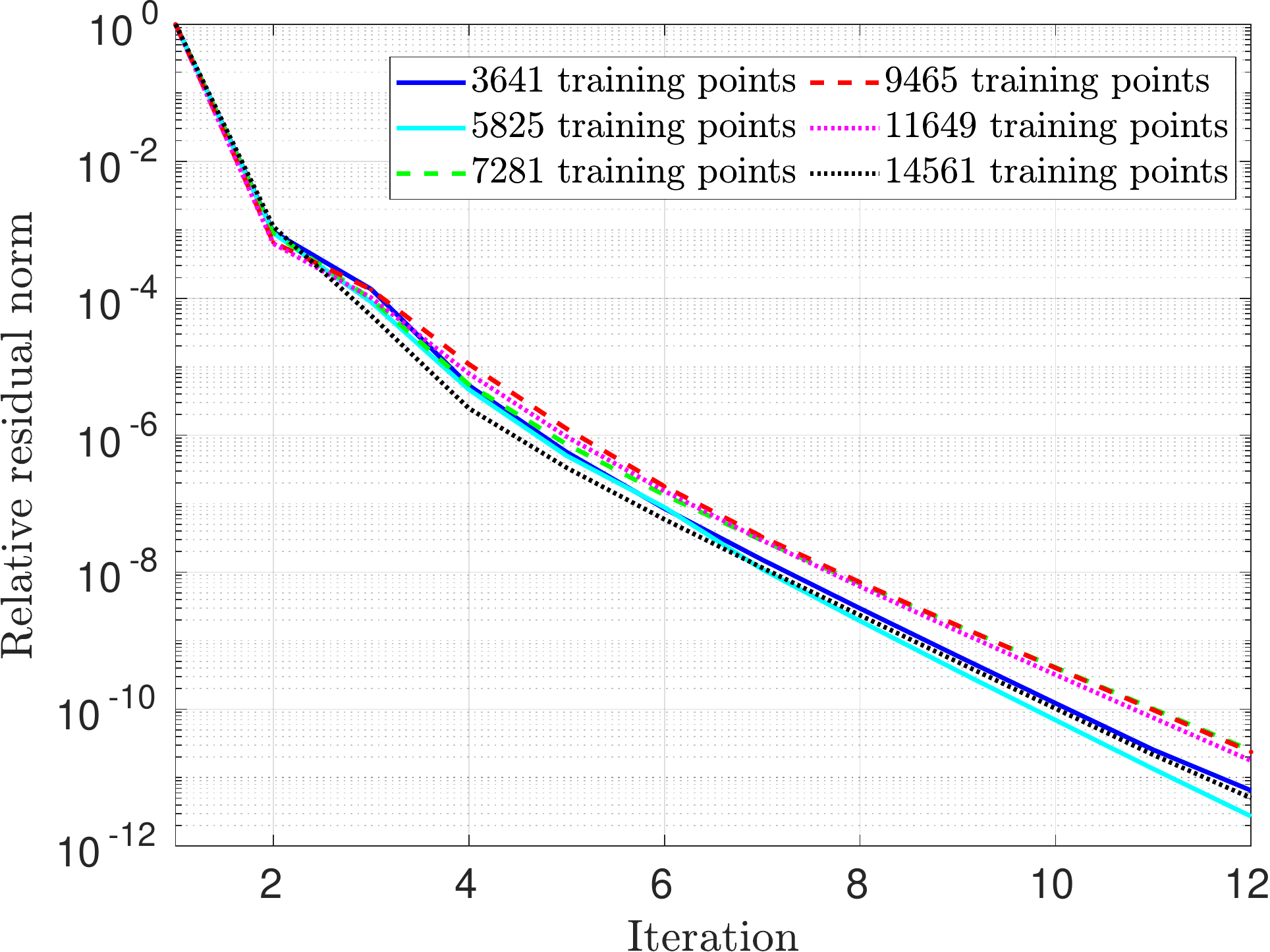}
\caption{laGPR - modified Newton-Raphson}\label{fig::Freeze}
\end{subfigure}
\caption{Convergence comparison over 12 Iterations for the benchmark clamped cube test obatined by adopting different machine learning techniques.}\label{fig:ConvergenceResults}
\end{figure}

Finally, the convergence behavior for the different machine learning techniques involving an increasing number of training points is studied in the context of a structural FE problem.
For this, a clamped cube with $8\times8\times8$ elements is loaded in three distinct cases, first in normal and in two shear directions (see Figure \ref{fig::ClampedCube}) such that a maximum absolute deformation gradient component value of $15\%$ from the undeformed configuration is reached. Then, the mean relative residual error norm $(L_{2})$ over the load cases is evaluated in one load step, over 12 nonlinear Newton-Raphson iterations  and 
visualized in
Figure \ref{fig:ConvergenceResults}. It can be seen that the naturally emerging discontinuities  of the KNN prediction of the material tangent are crucially preventing a proficient convergence behavior. These discontinuities are due to the fact KNN surrogates output the value of the nearest training point at a specific input. On the other hand, neural networks show a gradual but slow reduction of the residual norm.
Lastly Figures \ref{fig::noFreeze} and \ref{fig::Freeze} compare the laGPR convergence evolution when employing the introduced modified Newton-Raphson procedure. Due to the continuous updating of the surrogate model at each Gauss-point through every iteration, the approximation of the stress and material tangent outputs are constantly changing which does not allow the Newton-Raphson loop to convergence. However, by stopping the point-wise retraining and updating of the local GPR models once the Frobenius norm of the right Cauchy-Green residual reaches a value below $C_{tol}=0.01$ (that is, by adopting the modified Newton Raphson algorithm), a significant improvement in the convergence behavior compared to the original version and to neural networks is observed.

\newpage
\subsection{Homogenization of a high-fidelity microstructure}\label{sec::Homogenizatoin}
\begin{figure}[b!]
\begin{subfigure}[b]{.5\linewidth}
\centering
    \includegraphics[scale=0.2]{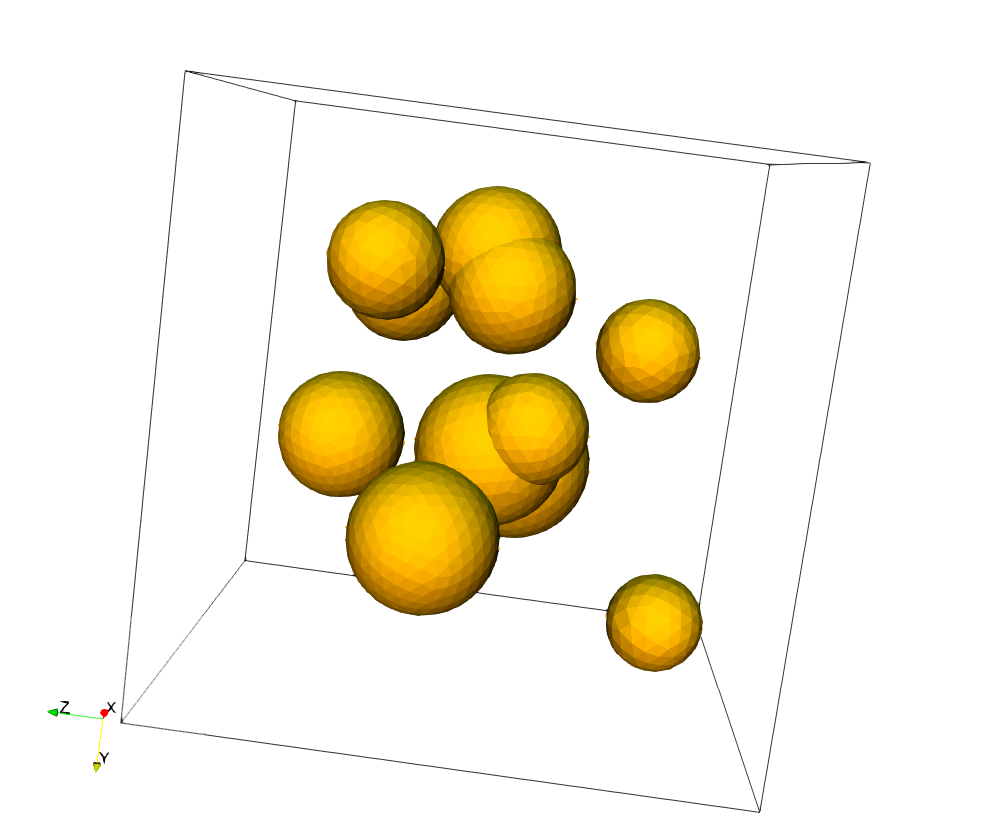}
\caption{}\label{}
\end{subfigure}%
\begin{subfigure}[b]{.5\linewidth}
\centering
    \includegraphics[scale=0.2]{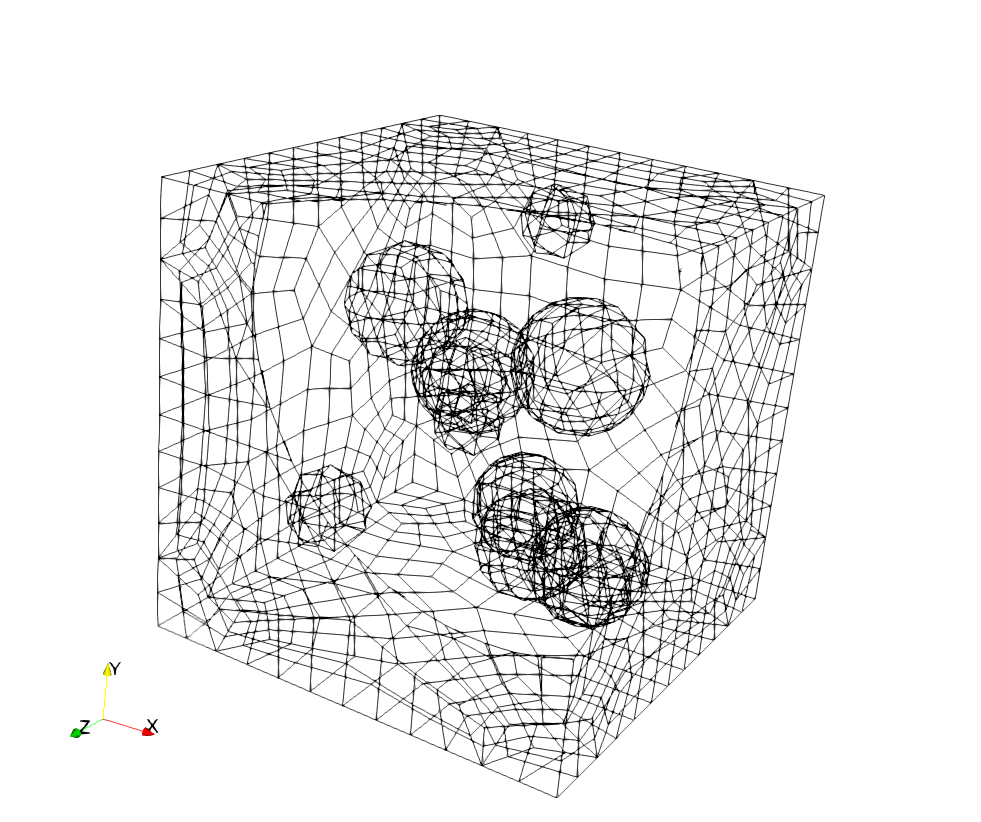}
\caption{}
\end{subfigure}
    \caption{Three dimensional RVE adopted in numerical simulations. (a) Twelve (in some cases partially overlapping) inclusions, are immersed in a homogeneous matrix (not shown), (b) Finite element mesh of the feature edges. }
    \label{fig:RVEHF}
\end{figure}

The local Gaussian-process regression approach is finally tested on data obtained from homogenization simulations of a microstructure with 12 inclusions, see Figure \ref{fig:RVEHF}.
We use a compressible Neo-Hookean formulation of the form
\begin{equation}
    \Psi = \frac{c_{1}}{\beta} (J^{-2 \beta} - 1) + c_{1} (I_{1} - 3)
\end{equation}
with
\begin{equation}
    \begin{aligned}
        \beta &= \frac{\nu}{1-2\nu}, \qquad c_{1} &= \frac{\mu}{2}.
    \end{aligned}
\end{equation}
The shear modulus in both phases is $80e3$ MPa.
The bulk modulus for the inclusions is $120e3$ MPa and for the matrix it is $160e3$ MPa respectively.

\begin{figure}[b!]
\begin{subfigure}[b]{1.0\linewidth}
\centering
\includegraphics[scale=0.18]{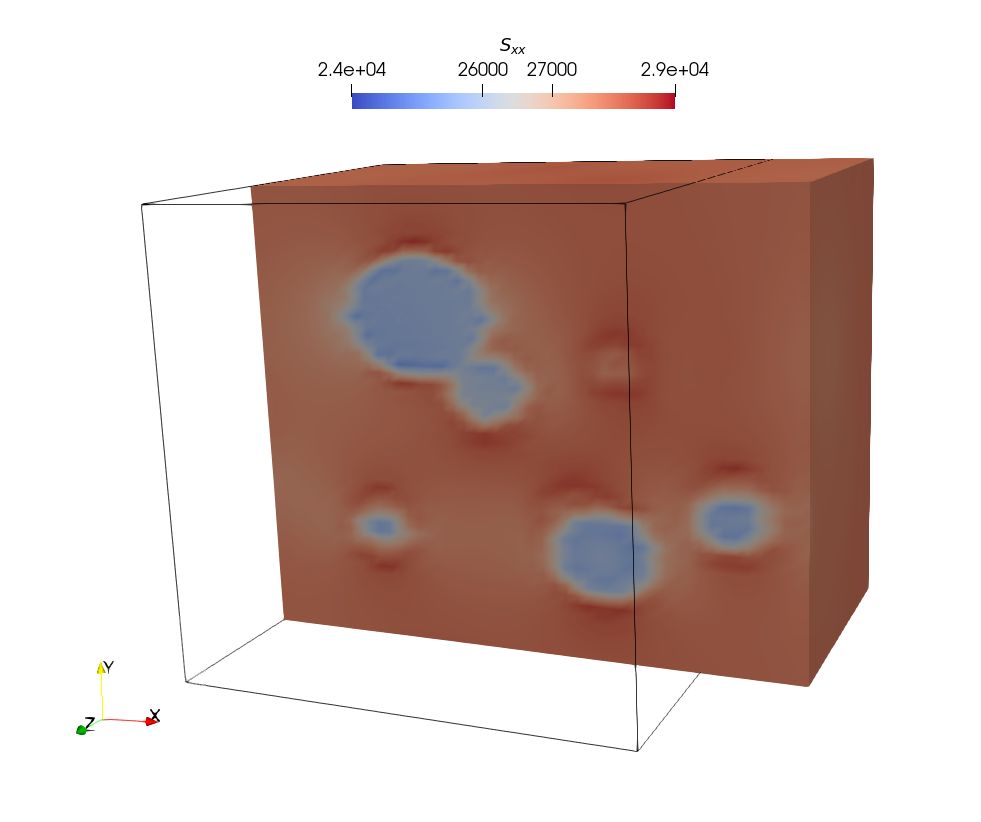}%
\includegraphics[scale=0.18]{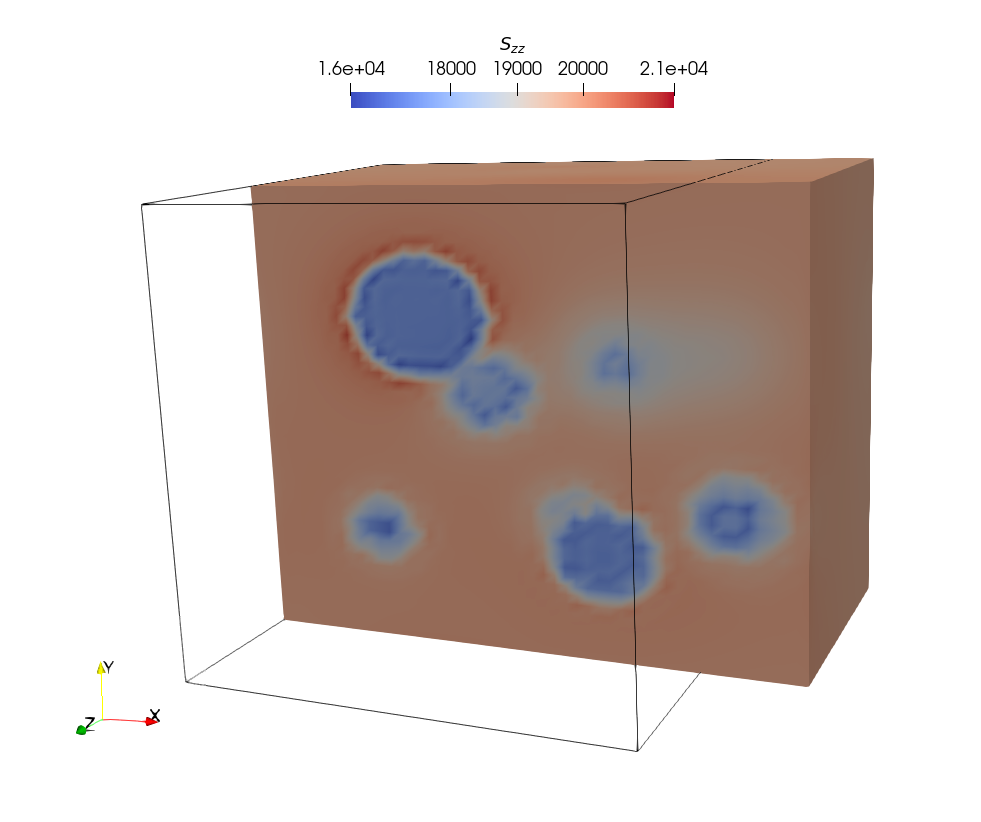}
\caption{$S_{xx}$ and $S_{zz}$ responses in uniaxial test}
\end{subfigure}

\begin{subfigure}[b]{1.0\linewidth}
\centering
\includegraphics[scale=0.18]{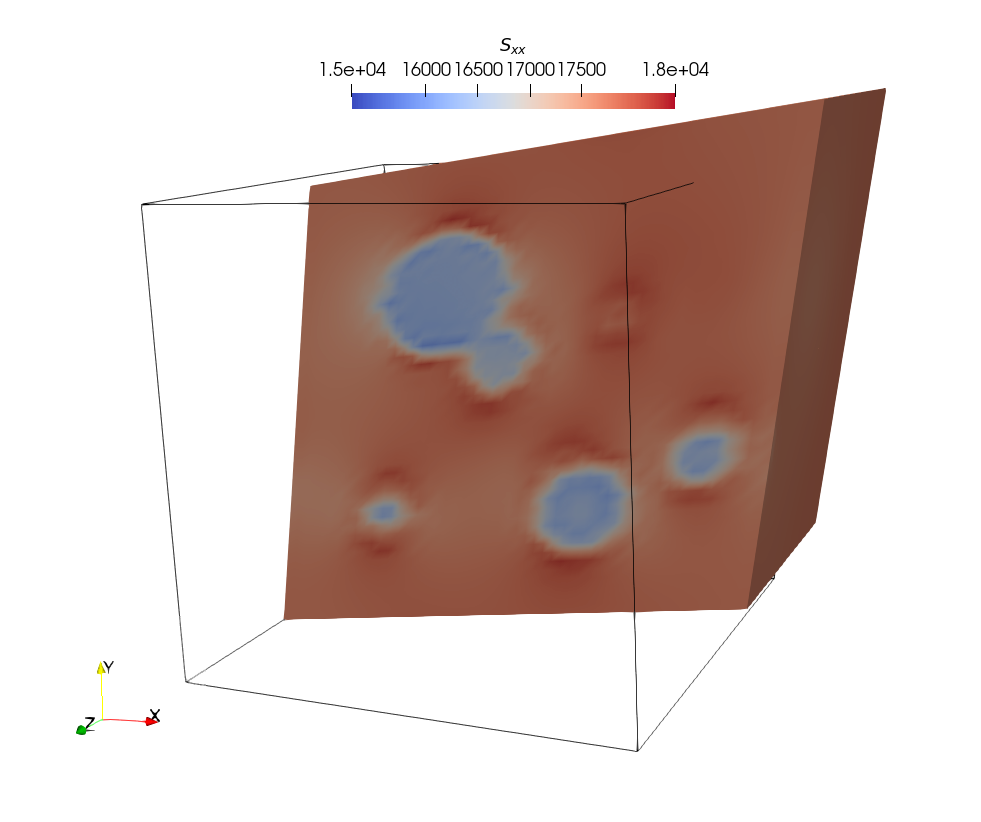}%
\includegraphics[scale=0.18]{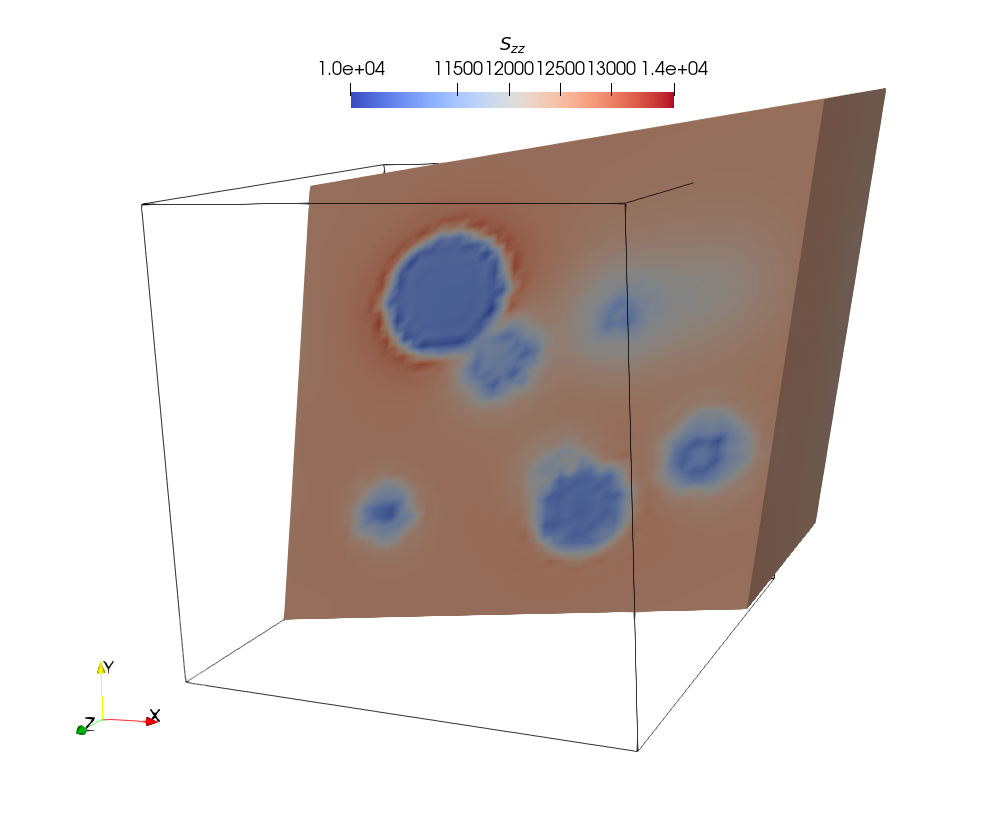}
\caption{$S_{xx}$ and $S_{zz}$ responses in multiaxial test}
\end{subfigure}
\caption{Microstructural response obtained with: (a)-(b) uniaxial displacement-driven test along x-direction, (c)-(d) mulitaxial displacement-driven test. Random cutting plane along z-direction. Black box represents undeformed RVE shape.}\label{fig::RVETest}
\end{figure}

\begin{figure}[b!]
\begin{subfigure}[b]{.5\linewidth}
\centering
\includegraphics[scale=0.8]{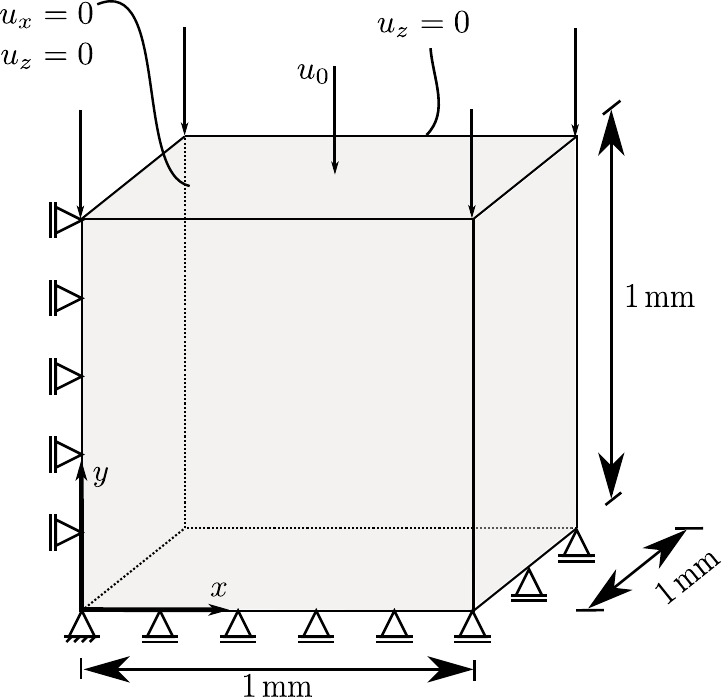} 
\caption{Punch test}\label{fig::PunTest}
\end{subfigure}%
\begin{subfigure}[b]{.5\linewidth}
\centering
\includegraphics[scale=0.8]{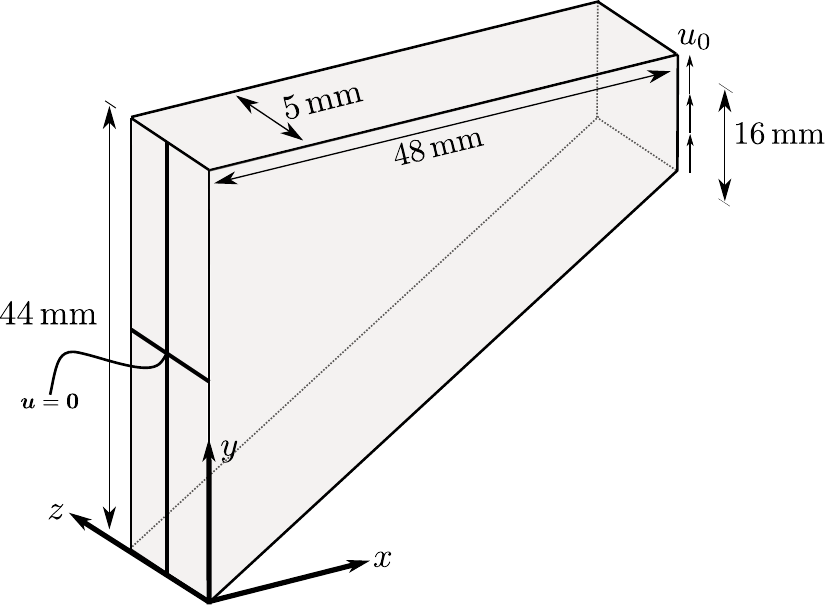}
\caption{Cook's membrane}\label{fig::Cook}
\end{subfigure}
\caption{Three dimensional macroscopic problems associated with the microscale heterogeneous RVE introduced in Figure \ref{fig:RVEHF}.}\label{fig::problems}
\end{figure}

\begin{figure}[b!]
\begin{subfigure}[b]{.5\linewidth}
\centering
\includegraphics[scale=0.2]{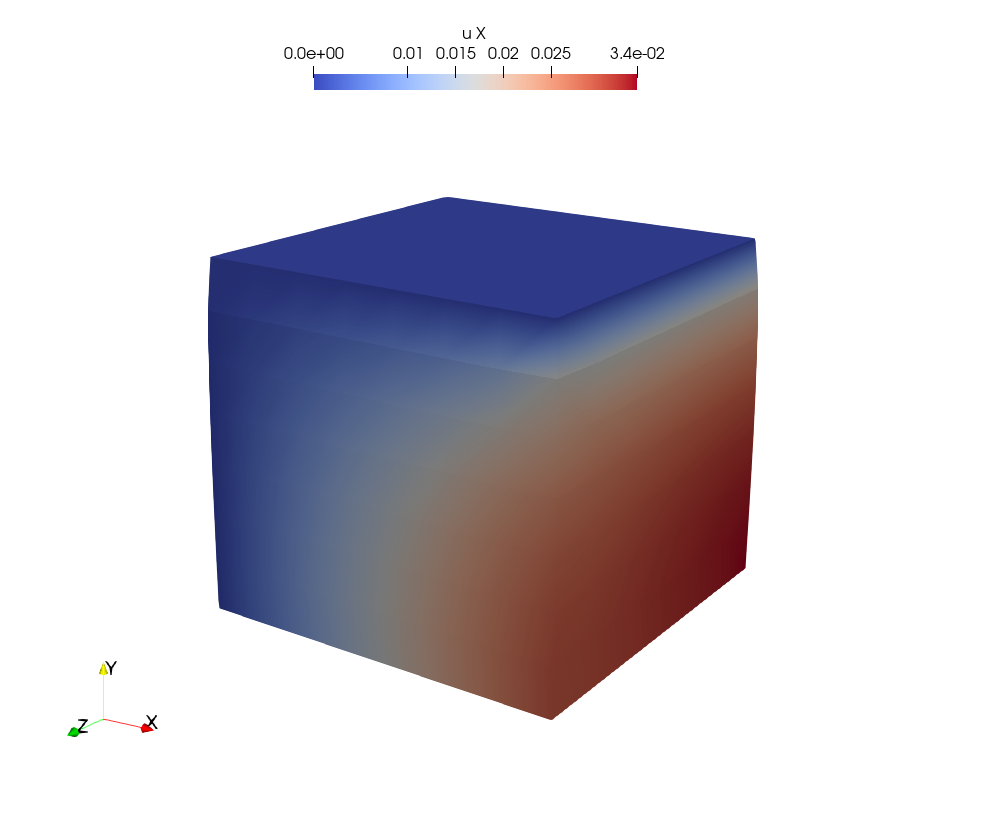} 
\caption{Punch test $u_x$}\label{fig::Punchux}
\end{subfigure}%
\begin{subfigure}[b]{.5\linewidth}
\centering
\includegraphics[scale=0.2]{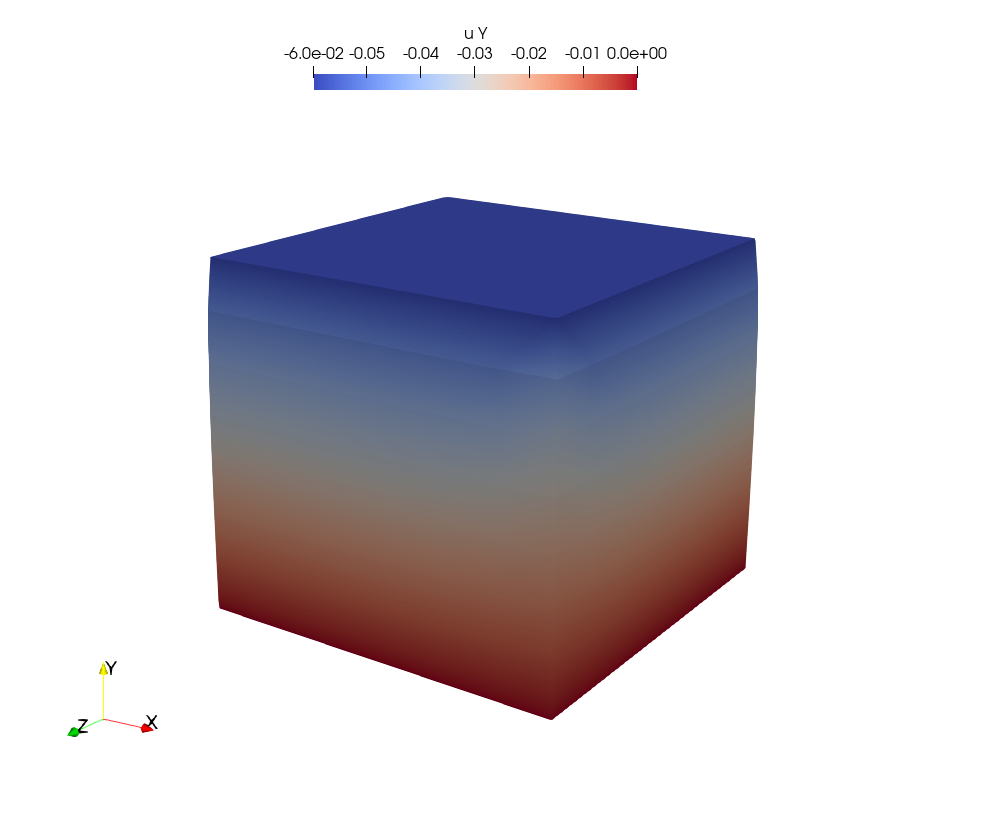}
\caption{Punch test $u_y$}\label{fig::Punchuy}
\end{subfigure}

\begin{subfigure}[b]{1.0\linewidth}
\centering
\includegraphics[scale=0.2]{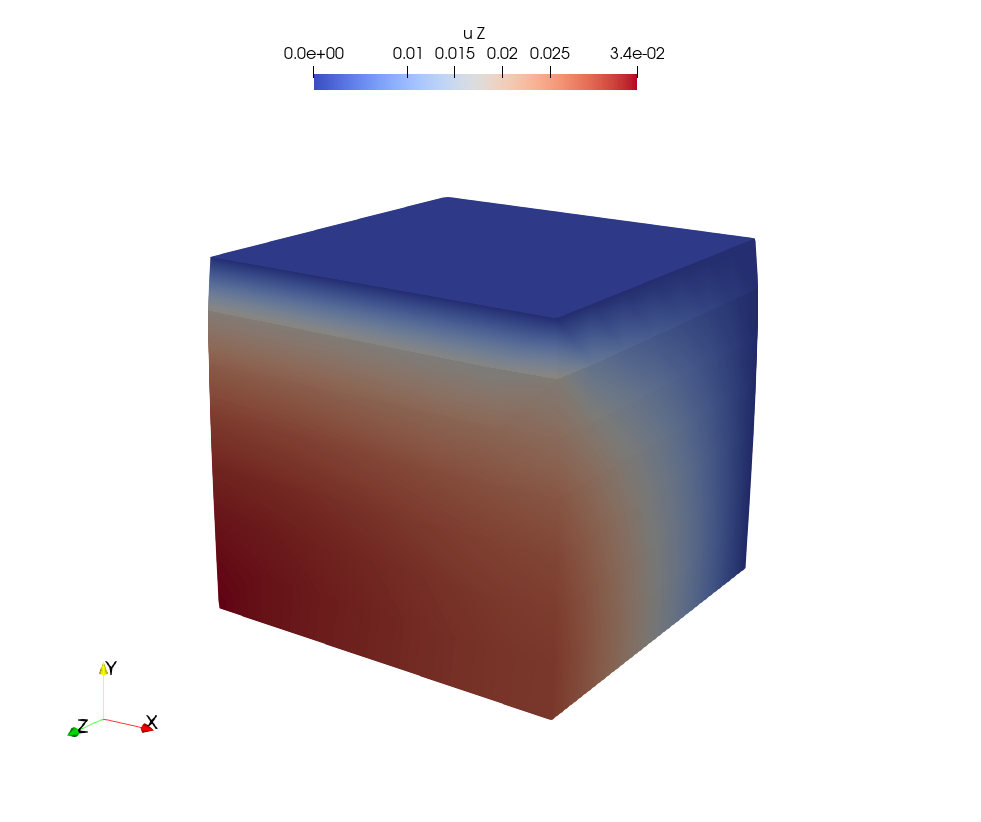}
\caption{Punch test $u_z$}\label{fig::Punchuz}

\begin{subfigure}[b]{.5\linewidth}
\centering
\includegraphics[scale=0.2]{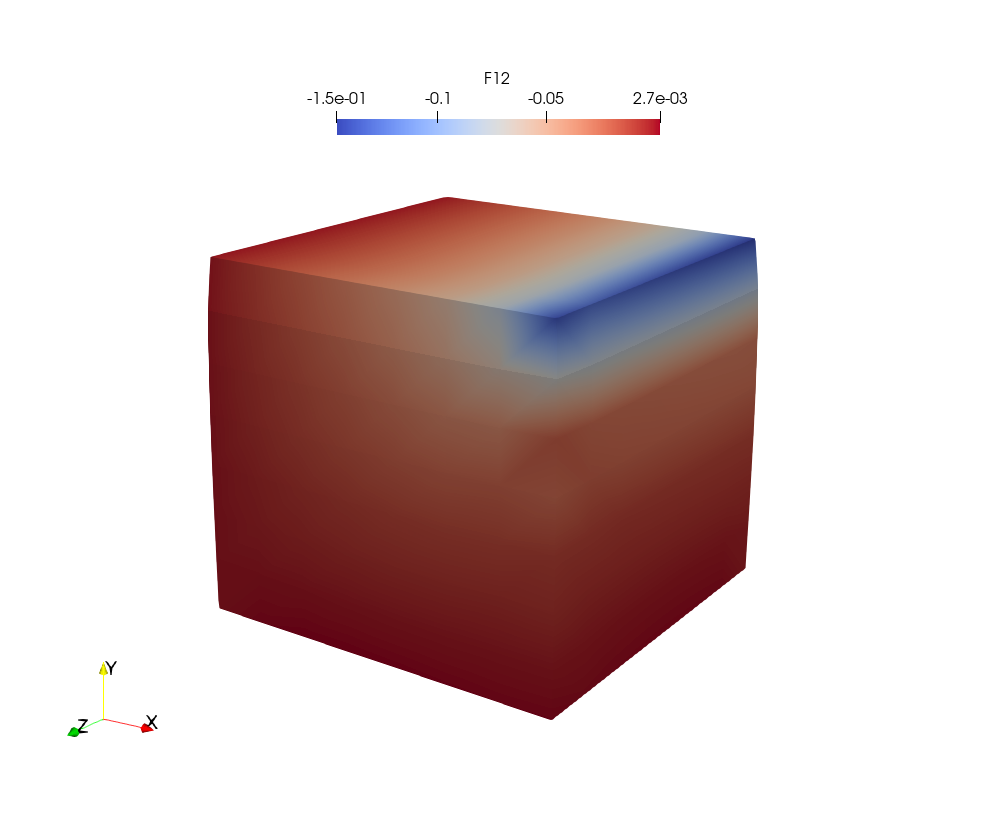} 
\caption{Punch test $F_{12}$}\label{fig::PunchF}
\end{subfigure}%
\begin{subfigure}[b]{.5\linewidth}
\centering
\includegraphics[scale=0.2]{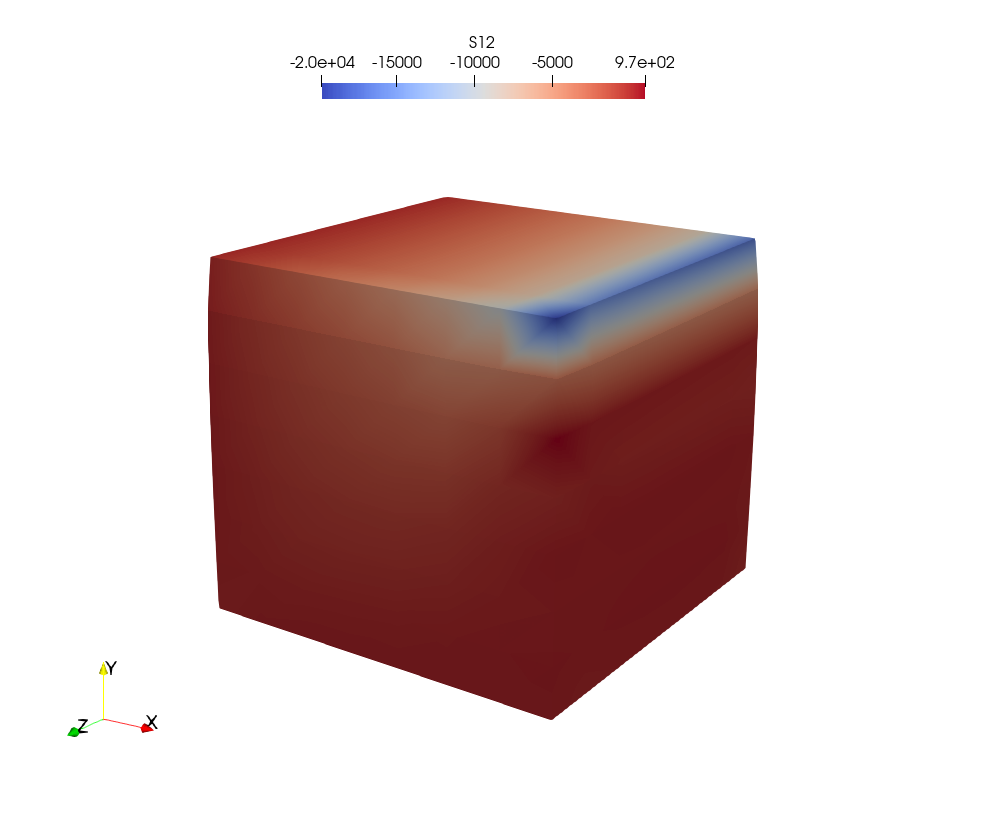}
\caption{Punch test $S_{12}$}\label{fig::PunchS}
\end{subfigure}
\end{subfigure}
\caption{Punch test results for $7\times 7 \times 7$ elements, maximum absolute strain $F_{12}=0.1491$ }\label{fig::ResponsesPunch}
\end{figure}

\begin{figure}[b!]
\begin{subfigure}[b]{.5\linewidth}
\centering
\includegraphics[scale=0.3]{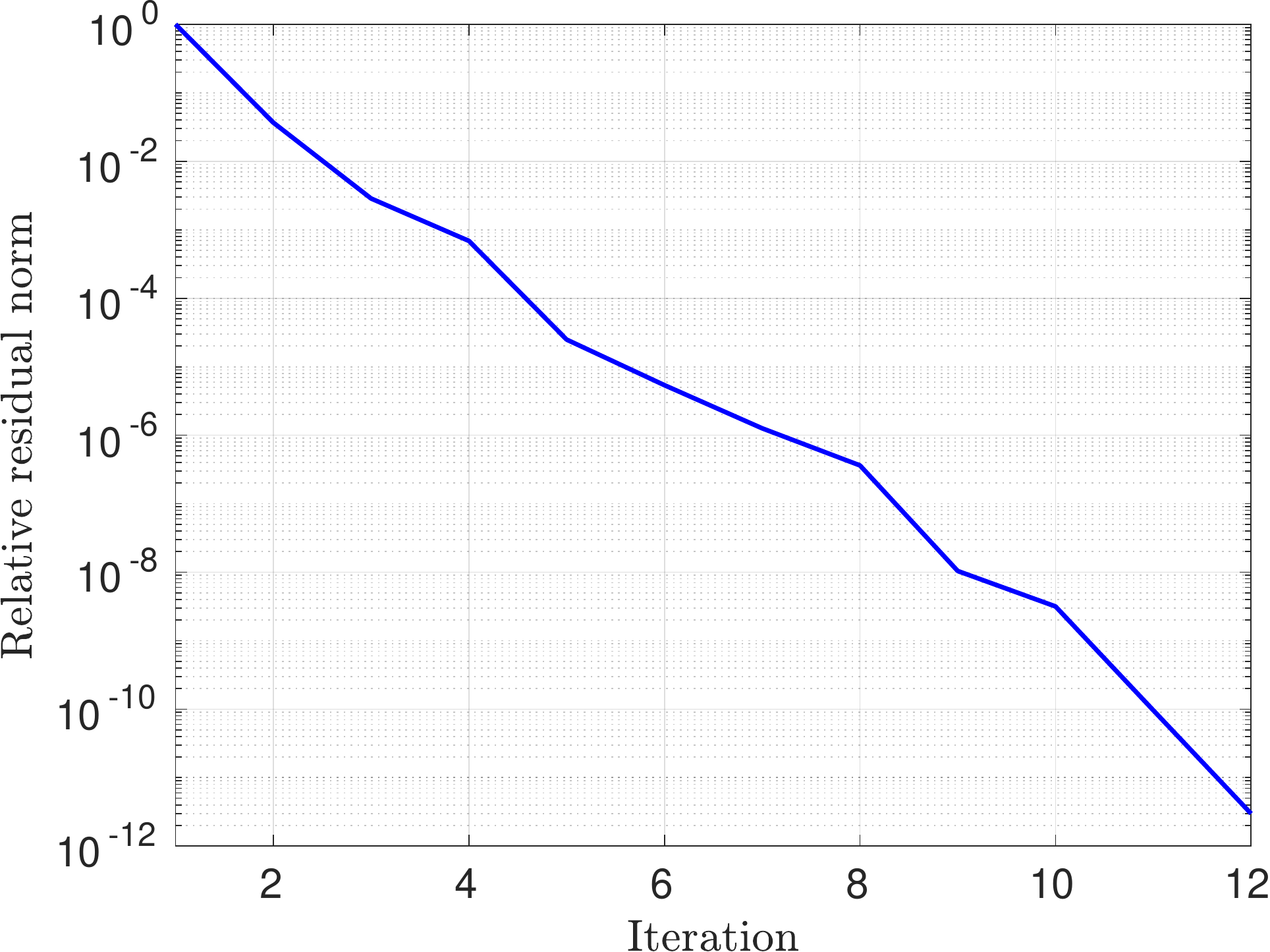} 
\caption{Punch test}\label{fig::PunTestConv}
\end{subfigure}%
\begin{subfigure}[b]{.5\linewidth}
\centering
\includegraphics[scale=0.3]{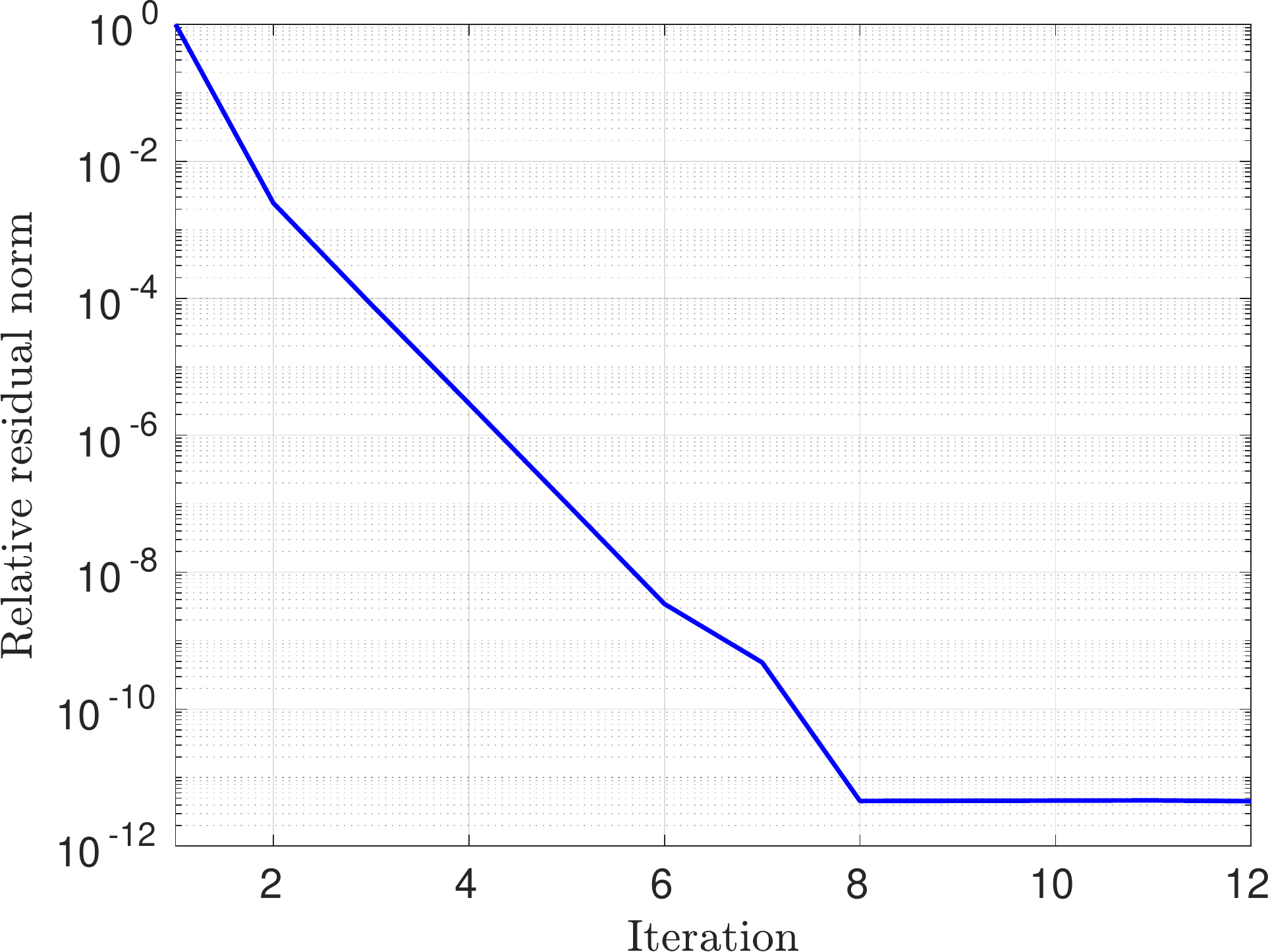}
\caption{Cook's membrane}\label{fig::CookConv}
\end{subfigure}
\caption{Relative residuals over nonlinear iterations for laGPR with 14561 training points.}\label{}
\end{figure}

\begin{figure}[b!]
\begin{subfigure}[b]{.5\linewidth}
\centering
\includegraphics[scale=0.2]{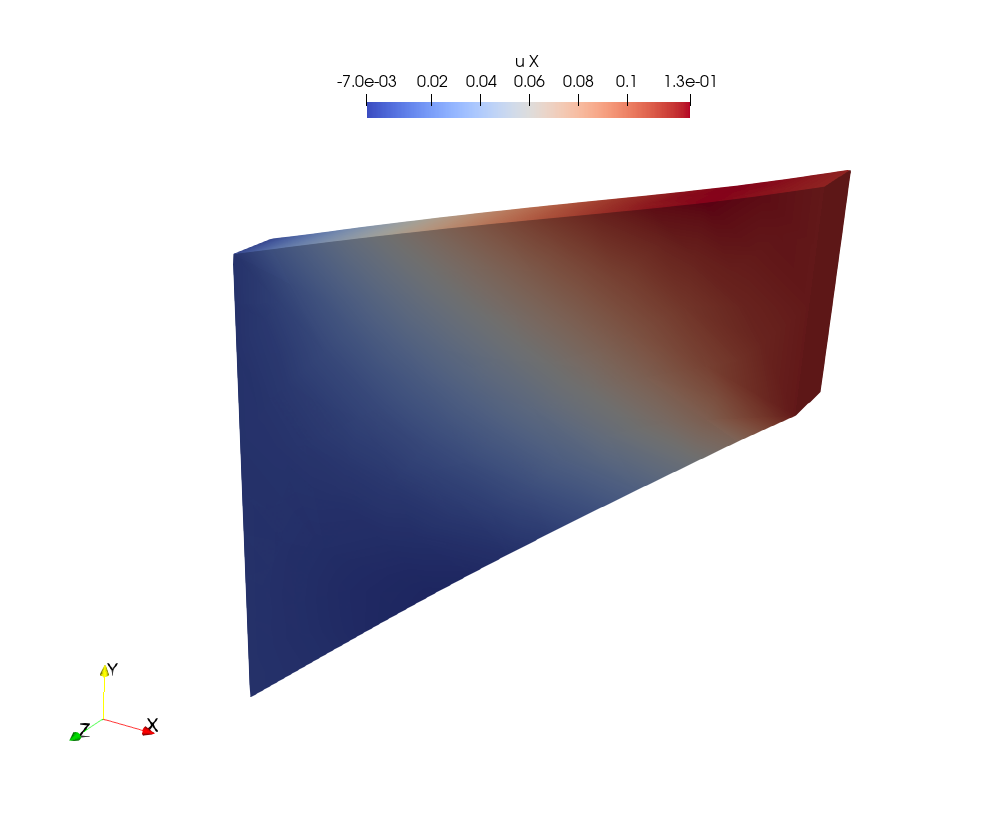} 
\caption{Cook's membrane $u_x$}\label{fig::Cookux}
\end{subfigure}%
\begin{subfigure}[b]{.5\linewidth}
\centering
\includegraphics[scale=0.2]{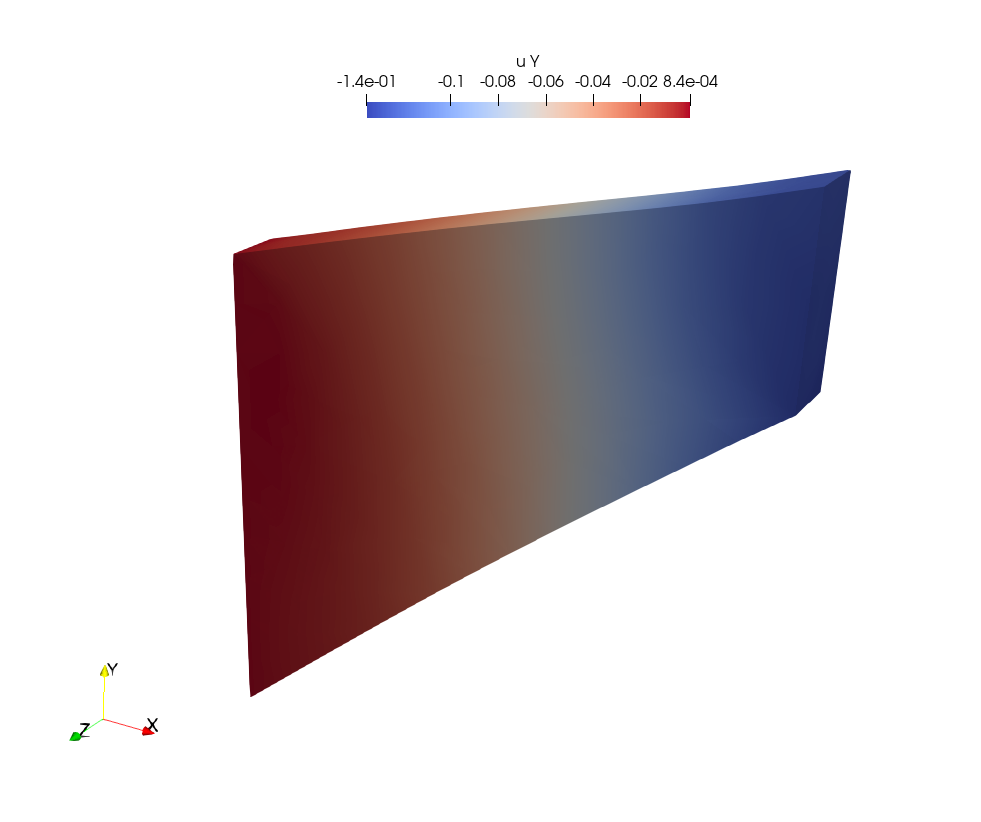}
\caption{Cook's membrane $u_y$}\label{fig::Cookuy}
\end{subfigure}

\begin{subfigure}[b]{0.5\linewidth}
\centering
\includegraphics[scale=0.2]{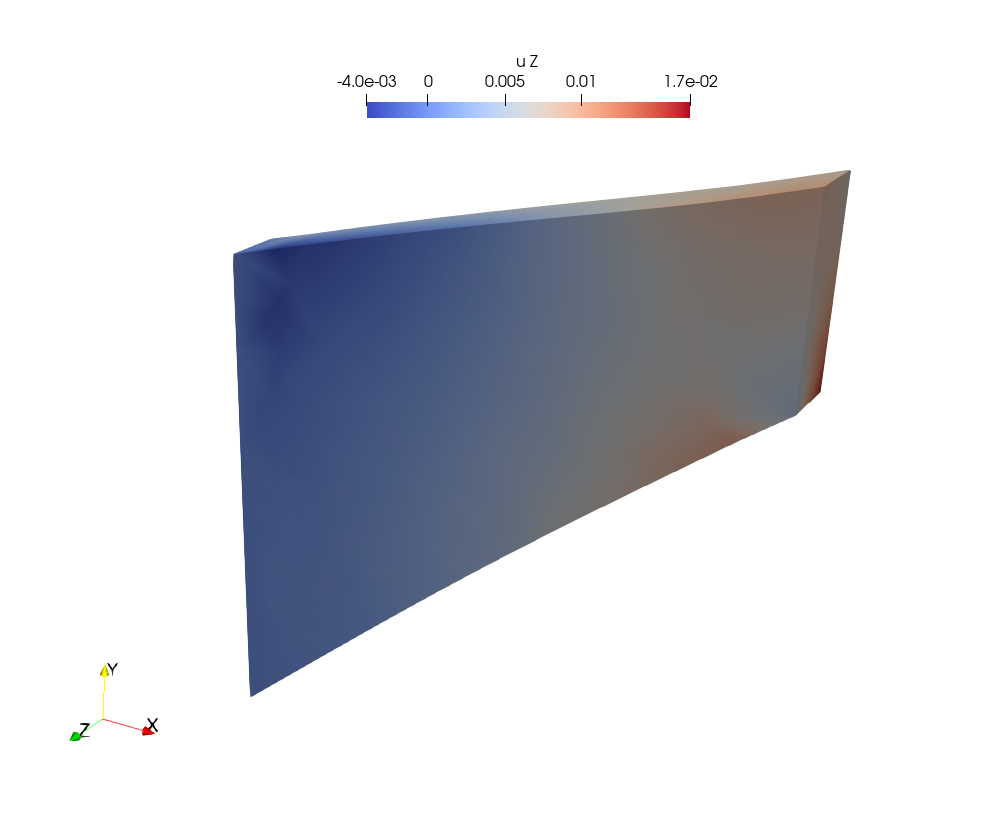}
\caption{Cook's membrane $u_z$}\label{fig::Cookuz}
\end{subfigure}%
\begin{subfigure}[b]{.5\linewidth}
\centering
\includegraphics[scale=0.2]{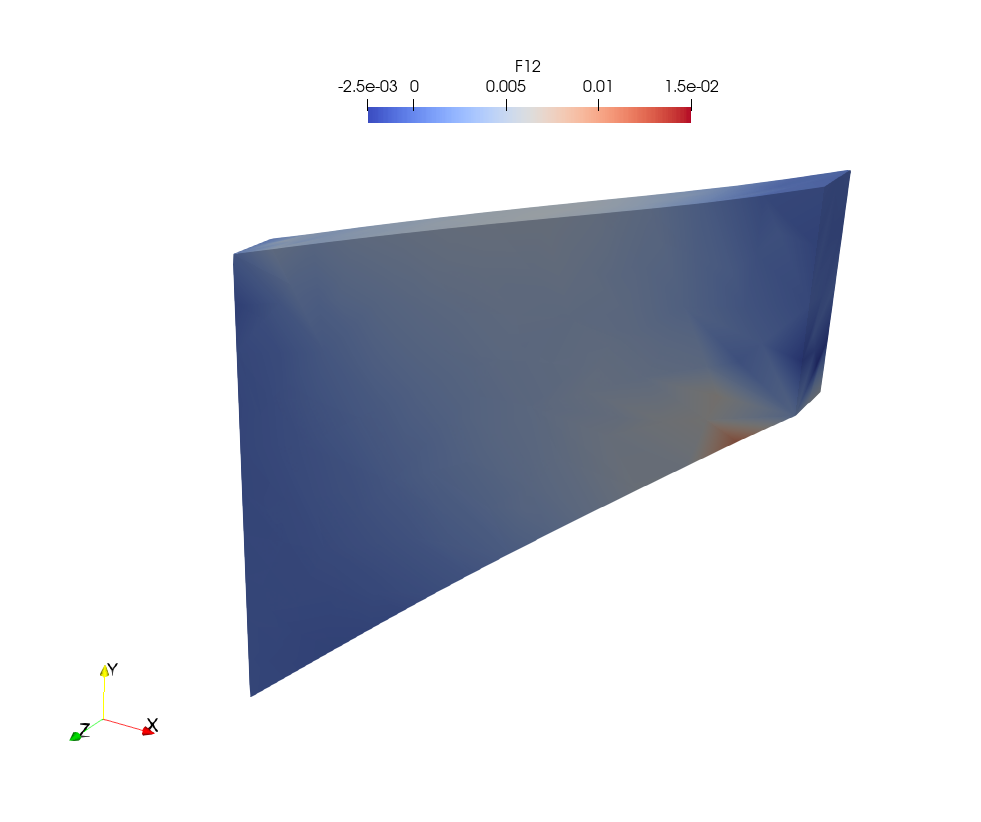} 
\caption{Cook's membrane $F_{12}$}\label{fig::CookF12}
\end{subfigure}%
\caption{Cook's membrane results, maximum absolute strain $F_{12}=0.1485$ }\label{fig::ResponsesCook}
\end{figure}

We generate 14561 input data combinations which is equivalent to 20 layers as described in Section \ref{sec::trainingPoints} in a $17.5\%$ training domain. The homogenization simulations were performed in C$++$ using the deal.ii-framework \citep{dealII92} with an author-modified version of the code provided by \cite{yaghoobi2019prisms}.
The RVE consists of around $25,000$ hexahedral elements. Figure \ref{fig::RVETest} displays responses to two sample microstructure loading cases, in particular a uniaxial and a multiaxial displacement-based test. One computation, which includes obtaining the consistent material tangent values following the approach of \cite{miehe1996numerical}, took around 8 minutes on a lab-cluster (2x AMD EPYC 7551 32C/64T, 16$\times$32Gb). Therefore running the data-generation code in parallel resulted in around 20 days of computing time. 
This observation highlights why a data-driven prediction of the microscale is needed when complex loading states are expected in the structural problem. Consider a general three dimensional structural problem consisting of $\approx 9,000$ elements. Then, without any load-stepping, and 12 nonlinear iterations $\approx 850,000$ calls would be made to a computational homogenization simulation of the microstructure.
Hence, a classical FE$^{2}$-scheme (without any parallelization) would need around 13 years to conclude with the same computing resources and implementation, which is of course intractable. 
Even lowering the number of elements to around $550$ would take $50,000$ calls, corresponding to around $280$ days total computation time.
Therefore, although 20 days for the initial data generation might sound time-consuming, it is in stark contrast to what would otherwise be necessary. Moreover, while when running traditional FE$^{2}$ computations the information is "lost" after the computation, i.e. a fully new computation is needed for a different loading case, the data generation step above covers all possible loading cases in a training hypercube of $17.5\%$ strain. As shown by the convergence response in Figure \ref{fig::Freeze}, less training points can result to similar convergence. Furthermore, Figure \ref{fig:ErrorBetweenMethods} shows that the error of the laGPR approximation of a constitutive law can be acceptable for lower number of training points, the offline cost can be suppressed significantly by choosing a smaller training dataset spanning the same convex hull.\\
After training, a Python-based laGPR script can be called at each Gauss point that trains the local Gaussian process model and provides the data-based predictions of the constitutive response. The interface was created employing the application programming interface offered by \citep{pythonCAPI}.
Depending on the number of inducing points of the laGPR algorithm, one constitutive law call takes between $0.2s - 1s$ on a conventional Laptop, speeding the macro-micro simulation up significantly.
\begin{figure}[b!]
\begin{subfigure}[b]{.5\linewidth}
\centering
\includegraphics[scale=0.4]{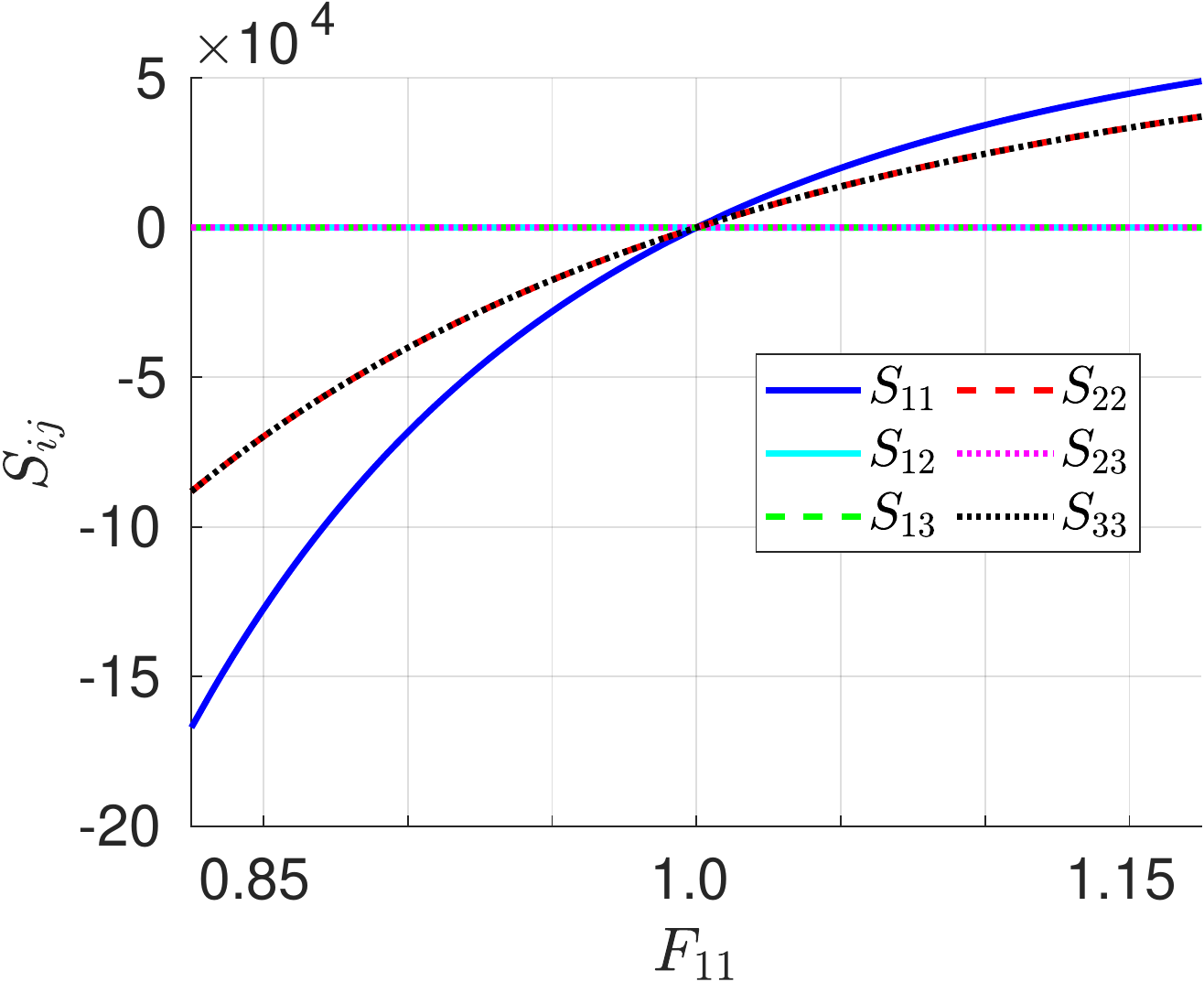} 
\caption{Stress over $F_{11}$}\label{}
\end{subfigure}%
\begin{subfigure}[b]{.5\linewidth}
\centering
\includegraphics[scale=0.4]{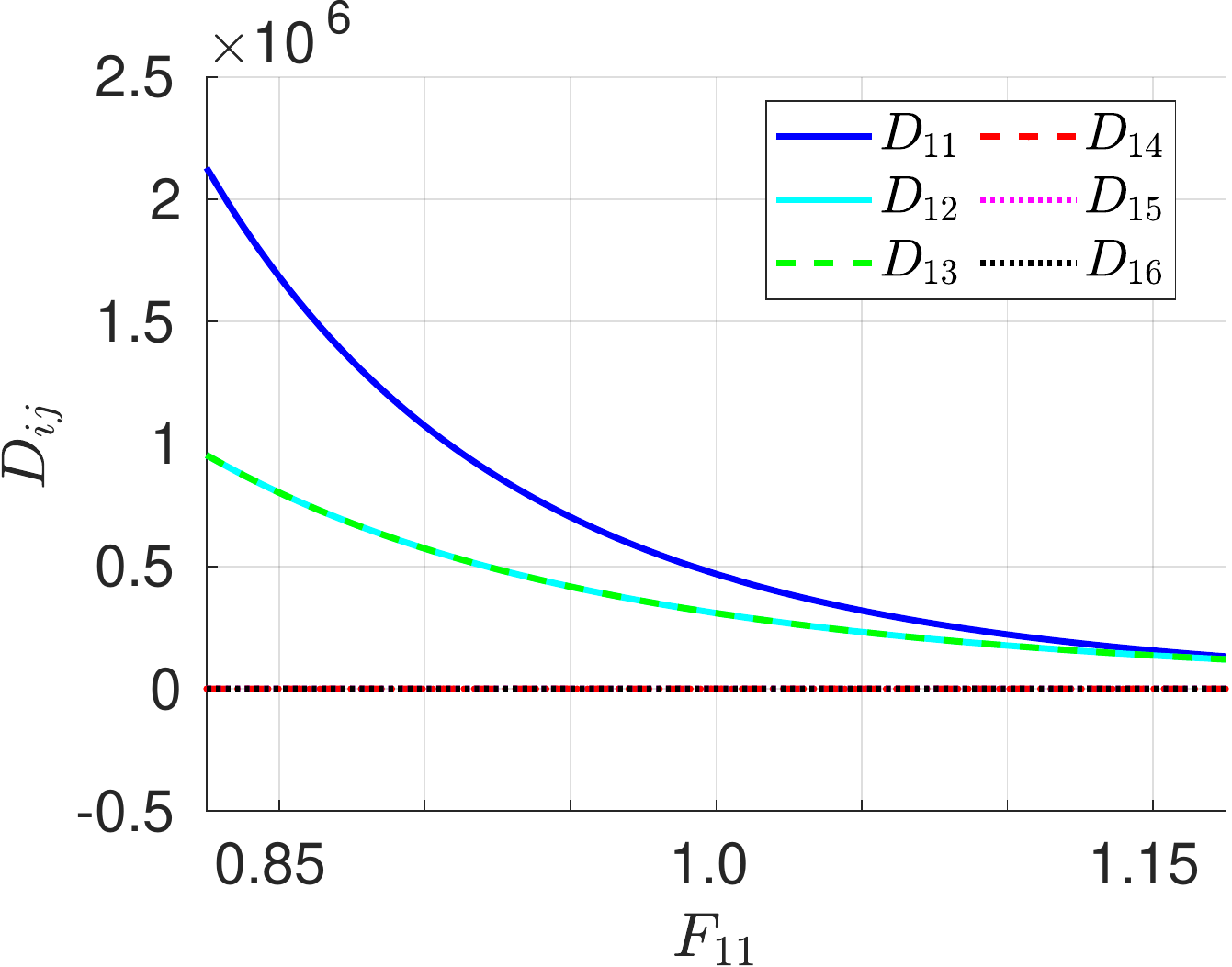}
\caption{Tangent over $F_{11}$}
\end{subfigure}%

\begin{subfigure}[b]{.5\linewidth}
\centering
\includegraphics[scale=0.4]{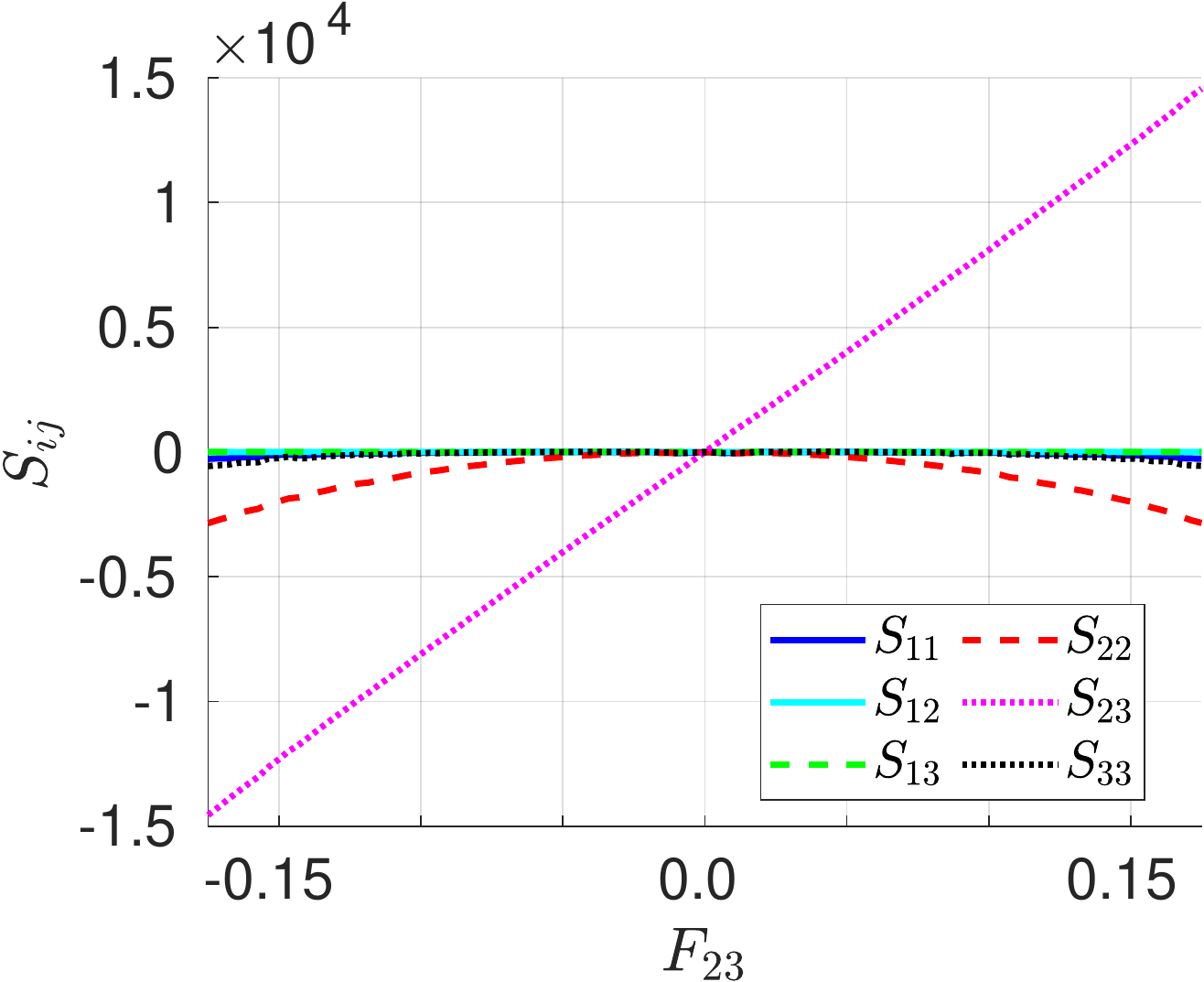} 
\caption{Stress over $F_{23}$}\label{}
\end{subfigure}%
\begin{subfigure}[b]{.5\linewidth}
\centering
\includegraphics[scale=0.4]{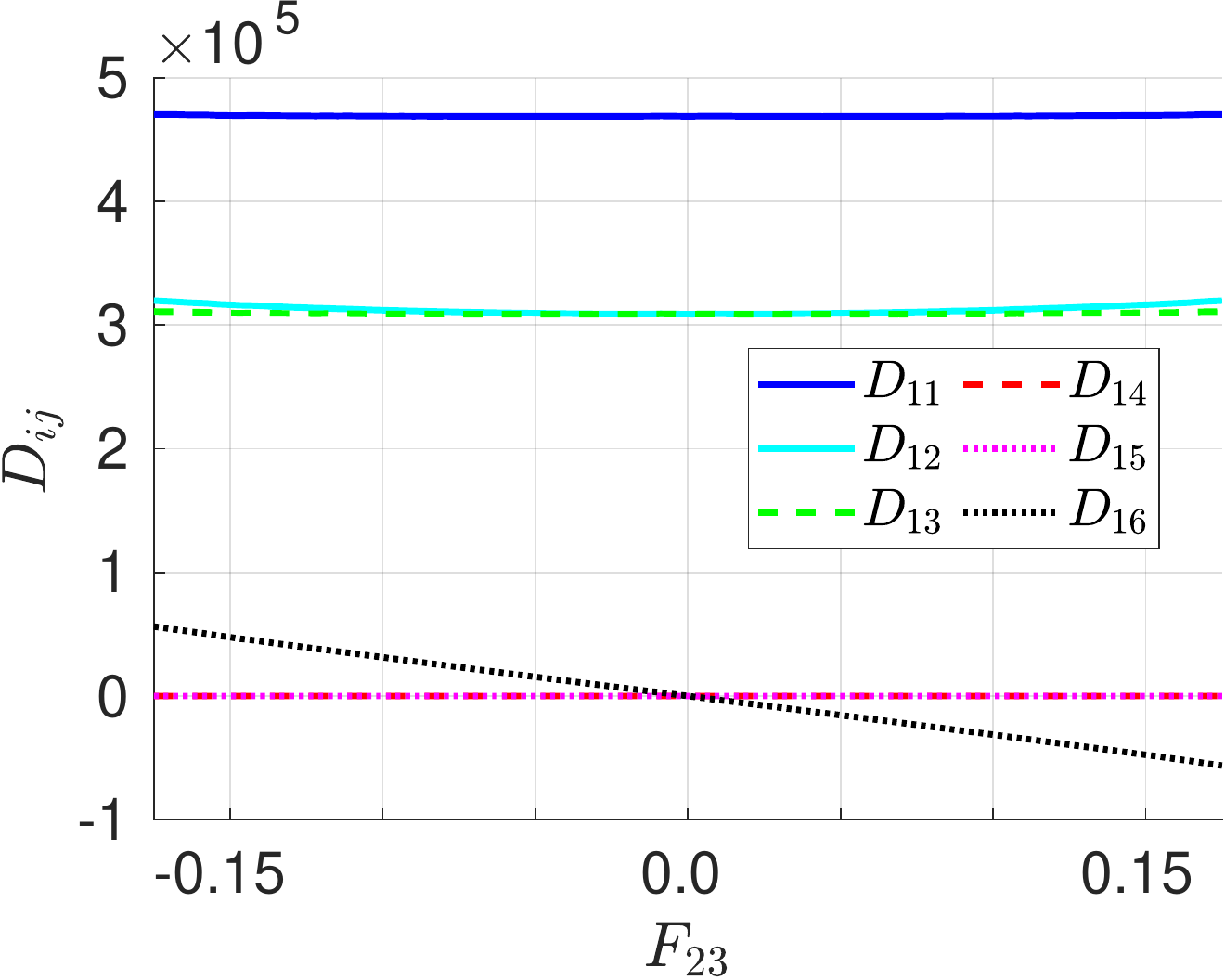}
\caption{Tangent over $F_{23}$}
\end{subfigure}
\caption{Fitted homogenized stresses and tangents of the heterogenouse RVE introduced in Figure \ref{fig:RVEHF} with variations from the undeformed configuration in the specified directions.}\label{fig::InclusionsConstitutiveResponse}
\end{figure}
The constitutive responses for two different displacement-based uniaxial load cases are shown in Figure \ref{fig::InclusionsConstitutiveResponse}.
We test the trained laGPR model on two standard structural applications: a punch-test problem described, in Figure \ref{fig::PunTest}, with $u_{0}=-0.06m$; a Cook's membrane, see Figure \ref{fig::Cook}, with $u_{0}=0.12m$. Both of these applications result in a maximum absolute deformation gradient component value of $15\%$ strain. Hence, they require a constitutive response on the upper limit of the investigated training hypercube region of $17.5\%$.
Some chosen results for the Punch test are displayed in Figure \ref{fig::ResponsesPunch}. The corresponding convergence evolution of the nonlinear Newton-Raphson procedure without any load-stepping is shown in Figure \ref{fig::PunTestConv}, which yields a similar relative residual value as the initial tests of Figure \ref{fig:ConvergenceResults}.
Some selected results for the Cook's membrane problem are shown in Figure \ref{fig::ResponsesCook}. The convergence behavior of this problem is highlighted in Figure \ref{fig::CookConv}.

\clearpage
\section{Discussion} \label{sec::Discussion}
From the results of Section \ref{sec::Comparison} it is evident that laGPR exhibits vastly superior approximation capabilities for the approximation of effective constitutive responses compared to ANNs. 
We see no reason to believe that this performance difference is only restricted to the investigated hyperelastic law and the specific training region but is indeed a generalizable phenomenon.
However, as pointed out in the introduction (Section \ref{sec::intro}), ANNs have evolved into the most commonly employed machine learning algorithm for constitutive response approximations and the results are generally satisfying. Nevertheless, the error convergences of ANNs with increasing datasets, as shown in Figure \ref{fig:ErrorBetweenMethods} are rarely reported and comparisons with other ML approaches are seldomly performed.

Additionally, in context of the structural Finite Element method, the convergence of the residual (an indication for the accuracy of the fit) are not investigated.
Indeed, we believe that if we had not performed these additional tests between the different methods for a variety of model hyperparameters we would have found the raw performances of neural networks to be satisfying as well, especially when using ANNs for problems where analytical solutions are not known and/or not enough test data is available. Additionally, the issues of accuracy of the major mapping might not be as prominent when trying to capture linear or weakly nonlinear responses in a 2D setting. \\Furthermore, ANNs appear to capture the constitutive behavior accurately enough when only looking at the responses along major uniaxial loading directions, see Figure \ref{fig::AnalyticalF11}. However, (at least) in context of three-dimensional hyperelastic laws they do not hold up when subjected to a closer look and detailed discussions, especially when compared to laGPR. In particular, the convergence of the relative residual norm of neural networks when used in the Finite-Element method, see Figure \ref{fig::ConvergenceANN}, points toward the fact that simply using backpropagation to obtain the material tangent does not yield optimal results and might need to be avoided in future works. Furthermore our empirical results do not find that adding a Sobolev term to the loss function of the ANNs helps to increase their performances. In fact, it appears that for the investigated datasets and the studied hyperparameters the complexity of the optimization problem got increased to a degree for which the accuracy of the final model was severely impaired.\\
We do not claim that this is a generalizable result and perhaps other hyperparameter settings or maybe other optimization schemes (other than ADAM) might have prevented these issues. However, we believe that this result very accurately highlights the problems associated with the parametricity of ANNs. In particular, because the authors already have extensive experiences with neural networks and were still not able to find a parameter setting that worked proficiently.
On the other hand, in Section \ref{sec::Homogenizatoin} we showed that laGPR can be applied to efficiently learn the constitutive responses of high-fidelity data obtained from computational homogenizations of a heterogeneous RVE. There was no required hyperparameter search and we solved the optimization problem (for the trainable parameters) with a standard optimization method.
Furthermore, it was shown that the convergence behavior (in terms of relative residual norm) that was achieved in the analytical benchmark case, carries over to the high-fidelity dataset. This proves the usefulness of the introduced modified Newton-Raphson algorithm.

\section{Conclusion}\label{sec::Conclusion}
This work describes a framework for data-driven constitutive modeling using local approximate Gaussian process regression for finite strain three dimensional hyperelastic laws, and proposes an extension for the utilization of these constitutive laws in structural FE problems. The suggested methodology is suggested in the context of computational homogenization and for the direct utilization of experimental data for constitutive modeling and FE simulations. It was motivated by the observation that problems arise when using data-trained artificial neural networks as a replacement for traditional constitutive laws in a nonlinear (finite-strain) framework, such as their parametric nature as well as unreliable convergence behavior. \\
The idea behind laGPR is that they approximate an output of a functional mapping by using only a small subset of the whole dataset, and this subset is in turn fitted by a standard Gaussian process regression model. In this work, the subset of inducing points is chosen based on a nearest neighbor approach. 
This work firstly compares laGPR and neural network using data originating from a benchmark analytical transversally isotropic constitutive formulation. Two different neural network training routine were tested, i.e. with and without employing derivative terms in the loss function. It was found that Sobolev training does not enhance the neural network quality in context of the investigated (low data) datasets.
However, it was found that laGPR is able to train constitutive models more accurately as well as in a more straightforward way compared to artificial neural networks.
In a second application, the method was tested in a high-fidelity environment by using data from computational homogenization simulations of a heterogeneous RVE as the training set. In order to aid the convergence of the method, a modified Newton-Raphson procedure was introduced. \\
We envision that this method can replace the classically used neural network formulations for one-to-one mapped data-driven constitutive models entirely. \\
In future work we aim to extend this method to time- and path dependent constitutive laws and model-data-driven approaches such as introduced in \cite{fuhg2021model}, as well as utilize experimentally generated data.

\clearpage
\typeout{}
\bibliography{bib.bib}

\end{document}